\documentclass[doc,apacite]{apa6}\usepackage[]{graphicx}\usepackage[]{color}

\makeatletter
\def\maxwidth{ %
  \ifdim\Gin@nat@width>\linewidth
    \linewidth
  \else
    \Gin@nat@width
  \fi
}
\makeatother

\newcommand{\tc}{\text{c}}
\newcommand{\tm}{\text{m}}
\newcommand{\R}{\text{r}}
\newcommand{\di}{\text{d}}
\newcommand{\D}{\text{D}}
\newcommand{\W}{\text{W}}
\newcommand{\E}{\text{E}}
\newcommand{\var}{\text{Var}}

\usepackage{apacite}
\usepackage{alltt}
\usepackage{epsfig,amsmath,alltt,bm,url}
\usepackage[english]{babel}

\IfFileExists{upquote.sty}{\usepackage{upquote}}{}

\title{Bayesian comparison of latent variable models: Conditional vs marginal likelihoods}
\twoauthors{E. C. Merkle}{D. Furr and S. Rabe-Hesketh}
\twoaffiliations{University of Missouri}{University of California, Berkeley}

\abstract{
Typical Bayesian methods for models with latent variables (or random effects) involve directly sampling the latent variables along with the model parameters. In high-level software code for model definitions (using, e.g., BUGS, JAGS, Stan), the likelihood is therefore specified as conditional on the latent variables. This can lead researchers to perform model comparisons via conditional likelihoods, where the latent variables are considered model parameters. In other settings, however, typical model comparisons involve marginal likelihoods where the latent variables are integrated out. This distinction is often overlooked despite the fact that it can have a large impact on the comparisons of interest. In this paper, we clarify and illustrate these issues, focusing on the comparison of conditional and marginal Deviance Information Criteria (DICs) and Watanabe-Akaike Information Criteria (WAICs) in psychometric modeling. The conditional/marginal distinction corresponds to whether the model should be predictive for the clusters that are in the data or for new clusters (where ``clusters'' typically correspond to higher-level units like people or schools). Correspondingly, we show that marginal WAIC corresponds to leave-one-cluster out (LOcO) cross-validation, whereas conditional WAIC corresponds to leave-one-unit out (LOuO). These results lead to recommendations on the general application of the criteria to models with latent variables.}

\keywords{
Bayesian information criteria,  conditional likelihood, cross-validation, DIC, IRT, leave-one-cluster out, marginal likelihood, MCMC, SEM, WAIC.}

\authornote{Correspondence to Edgar Merkle, email: merklee@missouri.edu. The research reported here was supported by NSF grant 1460719 and by the Institute of Education Sciences, U.S.\ Department of Education, through Grant R305D140059. The authors thank Fr\'{e}d\'{e}ric Gosselin and three anonymous reviewers for comments that improved the paper. Code to replicate the results from this paper can be found at \url{http://semtools.r-forge.r-project.org/}.}

\shorttitle{Bayesian comparison of latent variable models}

\begin{document}
\maketitle

\section{Introduction}

Psychometric models typically include latent variables (called ``random effects'' in some cases), classic examples being item response theory (IRT) models, structural equation models (SEMs), and multilevel/hierarchical regression models (MLMs). In IRT
and SEM, the latent variables represent latent traits or characteristics of {\em persons}, and in MLMs, they represent  random intercepts and possibly random slopes associated with {\em clusters}, such as schools, countries or, in the case of longitudinal data, persons. We will use the generic term ``clusters'' for the entities represented by latent variables in the model. In a Bayesian setting, latent variable models are hierarchical Bayesian models with latent variables as ``direct'' parameters, whose priors depend on hyperparameters.
Bayesian approaches have recently received increased interest due to improvements in the ease of coding and estimating complex models  by Markov chain Monte Carlo (MCMC) in various software packages. When it is computationally demanding to evaluate the \textit{marginal} likelihood (over the latent variables) for likelihood-based inference, MCMC has the great advantage of working with the \textit{conditional} likelihood (given the latent variables) by sampling the latent variables in tandem with the other model parameters. 

Comparison of Bayesian psychometric models is often facilitated by predictive information criteria, including the Deviance Information Criterion \cite<DIC; >{spibes02} and the Watanabe-Akaike Information Criterion \cite<WAIC; >{wat10}. These are readily computed by popular Bayesian software packages, and their computations are heavily based on the model likelihoods.
The sampling of latent variables can lead to a decision point related to the computation of the information criteria.
The decision point specifically involves the question of whether or not the latent variables should be treated as model parameters. If so, then we would compute information criteria via conditional likelihoods and expect ``effective number of parameter'' metrics to be large (each cluster has one or more unique latent variable values, and there are many clusters). If not, then we would compute information criteria via marginal likelihoods that are not generally available from the MCMC output. In traditional model estimation methods that optimize a likelihood or discrepancy function, the marginal likelihood (or some approximation of it) is used a large majority of the time.

Because DIC and WAIC are measures of models' abilities to make predictions for new units, the ``conditional vs.\ marginal'' issue can be framed as a question of the type of new data we wish to predict. If we want our models to make predictions for new units from the same clusters as in our original data, then the conditional likelihood (conditioning on the latent variables of the specific clusters observed) is appropriate.  Conversely, if we want our models to make predictions for new units from clusters not in the original data, then the marginal likelihood is appropriate.
We can also think of this distinction in terms of how we would define the folds in $k$-fold cross-validation and leave-one-out approaches, or in other words, what we would leave out during estimation in order to evaluate predictive accuracy. To generalize to new units from existing clusters, we would leave
out individual units from these clusters, and the predictive distribution for these units
would exploit information from other units in the same clusters by conditioning on the cluster-specific latent variable values. To generalize to new units from new clusters, we would leave out intact clusters, and the predictive distribution would be marginal over the latent variables associated with the clusters.
 From this standpoint, the marginal likelihood may generally be preferred in psychometric modeling; we typically wish to generalize inferences beyond the specific clusters that were observed in our data.
There are additional reasons to generally prefer the marginal likelihood, including the incidental parameter problem \cite{neyscot48,lan00}, of which a Bayesian version exists when Bayes modal inference is used \cite<also see>{mis86,oha76}.
 
\citeA{spibes02} discuss the conditional/marginal distinction in their original paper on the DIC, where they refer to the issue as ``model focus'' \cite<see also>{celeux2006deviance,mil09,tregel03}. If the parameters ``in focus'' include the latent variables, the likelihood becomes what we call the conditional likelihood; otherwise the likelihood is what we call the marginal likelihood. However, the distinction is hardly ever discussed in applications. Bayesian books targeted at applied psychometricians
 \cite<e.g.,>{song2012,kaplan2014,levy2016} and papers on Bayesian model comparison in psychometrics \cite<e.g.,>{kang2009,licohenetal2009,zhustone2012} define either the conditional
 DIC or a generic DIC for non-hierarchical models, then apply the conditional DIC without discussing the implications of this choice or mentioning the marginal DIC. \citeA{fox2010} is an exception here, where the generic DIC is first presented and various types of DICs (conditional and marginal) are later utilized in applications. \citeA{zhang2019} also discuss various forms of DIC in the context of multilevel IRT models.

 The WAIC has not yet been used much in psychometrics, exceptions being \citeA{luo2017} and \citeA{dasilva2018}, who
 define the conditional version of WAIC for IRT  without mentioning the marginal alternative, and \citeA{zhaohua2017} who use the marginal version in factor analysis, without discussing  the issue or mentioning the conditional alternative.
Moving beyond psychometrics, \citeA{gelhwa14} point out that there is a choice to be made when defining the likelihood for Bayesian information criteria (including WAIC) between what we call conditional and marginal versions, but they use only the conditional version of WAIC in their ``8-schools data'' application and comparison to leave-one-out approaches. 
 We are aware of a small number of 
 contributions that discuss the merits of the marginal version of WAIC~\cite{li2016approximating,millar2018}, but these treatments are confined to the special case of one response per latent variable (i.e., one unit per cluster). Further, \citeA{vehtari2016} define a marginal version of WAIC for Gaussian latent variable models without contrasting it to the conditional version.

From an applied perspective, researchers estimating Bayesian models are often unaware of the distinction between marginal and conditional versions of information criteria, relying on default software behavior. Users of BUGS \cite{bugs12,luntho00}, JAGS \cite{plu03}, and Stan \cite{stan14} typically obtain information criteria based on a conditional likelihood, whereas users of blavaan \cite{merros17} and Mplus \cite<e.g.,>{mutasp12} obtain information criteria based on a marginal likelihood.
Therefore, a first contribution of our paper is to emphasize the importance of the conditional/marginal distinction for information criteria and, subsequently, to recommend marginal criteria as a default.
We relatedly clarify other differences in the definition of the DIC implemented in various pieces of software.
 A second contribution is to propose a marginal version of WAIC for the prevalent case of multiple units per cluster
 and to contrast it with the conditional version. We specifically show that the marginal WAIC corresponds to leave-one-cluster out (LOcO)  cross-validation, whereas the conditional WAIC corresponds to leave-one-unit out (LOuO) cross-validation and has no rationale in the case of one unit per cluster.
 A third contribution is 
 to  propose an efficient version of adaptive quadrature for the case where the marginal likelihood does not have a closed form. The method differs from traditional adaptive quadrature because it is performed as a postprocessing step using MCMC samples of the model parameters and latent variables as input. Our idea of using the {\em marginal} (with respect to the model parameters) posterior first and second moments of the latent variables in the adaptive quadrature approximation can easily be applied in Monte Carlo integration by using the corresponding normal density as importance distribution, and this would be an improvement over the method proposed by \citeA{li2016approximating}.
 A fourth contribution is to provide two real examples, one using SEM and the other using IRT, where conditional and marginal information criteria can lead to very different substantive conclusions. For these examples, a large number of Monte Carlo draws is often needed to obtain acceptable Monte Carlo errors for the information criteria, especially in the conditional case. We  make some practical  recommendations for assessing convergence of the information criteria there.

In the next section, we define the class of models considered and in Section 3 we define and contrast information criteria based on conditional and marginal likelihoods. Two examples
based on real data are presented in Section 4, one using SEMs (Section 4.1) and the other IRTs (Section 4.2). We end with a discussion in Section 5.
 The supplementary material includes R code to estimate the IRTs and SEMs presented in this paper and obtain the information criteria -- by performing adaptive quadrature integration
 for IRT and by using  the recently extended R package blavaan for SEM.

\section{Models and conditional versus marginal likelihoods}
We consider models with continuous latent variables, such as IRTs, SEMs and MLMs. Responses $y_{ij}$ are observed for units $i$ belonging to clusters $j$, such as students $i$ in schools $j$ or items/indicators $i$ for persons $j$, whereas the latent variables $\bm{\zeta}_j$ are cluster-specific. The responses follow
a generalized linear (or similar) model in which the conditional expectation of the responses is a function of the latent variables $\bm{\zeta}_j$ and possibly observed covariates. The {\em conditional likelihood} is
\begin{equation}
f_\tc(\bm{y}| \bm{\omega}, \bm{\zeta})\ = \   \prod_{j=1}^J  f_\tc(\bm{y}_{j} | \bm{\omega}, \bm{\zeta}_j),
\label{eq:likec}
\end{equation}
where $\bm{y}_{j}$ is the vector of responses for the units in cluster $j$ ($j=1,\ldots,J$), $\bm{y}$ is the vector of responses for all units (the $\bm{y}_j$ stacked on one another),
$\bm{\zeta}_j$ is the vector of latent variables for cluster $j$,
$\bm{\zeta}$ is the vector of latent variables for all clusters (the $\bm{\zeta}_j$ stacked on one another),  and  $\bm{\omega}$ are model parameters. Under conditional independence, the joint conditional density (or probability)
of the responses for cluster $j$ factorizes as
\[f_\tc(\bm{y}_{j} | \bm{\omega}, \bm{\zeta}_j)\ = \ \prod_{i=1}^{n_j} f_\tc(y_{ij} | \bm{\omega}, \bm{\zeta}_j),\]
where $f_\tc(y_{ij} | \bm{\omega}, \bm{\zeta}_j)$ is the conditional
density  of the response for unit $i$ in cluster $j$ ($i=1,\ldots,n_j$), given $\bm{\omega}$, $\bm{\zeta}_j$, and possibly covariates (not shown).

For example, in a confirmatory factor analysis, $y_{ij}$ is a continuous response for indicator $i$ and person $j$, $\bm{\zeta}_j$ is a vector of common factors, $f_\tc(y_{ij} | \bm{\omega}, \bm{\zeta}_j)$ is a normal density, and $\bm{\omega}$ includes intercepts, factor loadings, and unique factor variances. In an item response model with binary items, $f(y_{ij} | \bm{\omega}, \bm{\zeta}_j)$ is $\pi_{ij}^{y_{ij}}(1-\pi_{ij})^{1-y_{ij}}$ where $\pi_{ij}$ is the
conditional probability that $y_{ij}$ equals 1 (often an inverse logit or inverse probit of a linear predictor), $\bm{\zeta}_j$ are dimensions of ability, and $\bm{\omega}$ includes difficulty parameters and possibly discrimination and guessing parameters. We call $f_\tc(\bm{y}| \bm{\omega}, \bm{\zeta})$  the conditional likelihood because it is conditional on the latent variables. Note that, in likelihood-based inference, ``conditional likelihood'' usually refers to the likelihood in which the latent variables
have been eliminated by conditioning on sufficient statistics, whereas inference based on our conditional likelihood above is sometimes referred to as joint maximum likelihood estimation.

The latent variables are independent across clusters and have a joint prior distribution \[g(\bm{\zeta}|\bm{\psi})=\prod_{j=1}^J g(\bm{\zeta}_j|\bm{\psi}_j)\]
with hyperparameters $\bm{\psi}$ that potentially include different values $\bm{\psi}_j$ for different
(groups of) clusters $j$.
Specification of the Bayesian model is completed by assuming prior distributions  for $\bm{\omega}$ and $\bm{\psi}$, typically
$p(\bm{\omega}, \bm{\psi})=p(\bm{\omega})p(\bm{\psi})$.

The {\em marginal likelihood} is
\begin{equation}
 f_\tm(\bm{y}| \bm{\omega}, \bm{\psi})\ = \   \prod_{j=1}^J \int f_\tc(\bm{y}_{j} | \bm{\omega}, \bm{\zeta}_j) g(\bm{\zeta}_j|\bm{\psi}_j) \di\bm{\zeta}_j.
 \label{eq:likem}
\end{equation}
We call this likelihood marginal because it is integrated over the latent variables. However, this likelihood is not marginal over the model parameters and should therefore not be confused with the marginal likelihood that is used, for instance, for Bayes factor computation. The marginal likelihood defined above is the standard likelihood employed in
maximum likelihood estimation of latent variable or random-effects models. Its use implies that the latent variables are not of direct inferential interest, instead being ancillary parameters that are removed from the likelihood.
\citeA{celeux2006deviance}, who regard the latent
variables as missing data, refer to this likelihood as the observed likelihood because it is marginal over the missing data.

Note that there are usually some parameters that can be included either in
$\bm{\psi}$ or in $\bm{\omega}$,  depending on how the model is formulated. For example, a linear two-level variance-components
model can be written
two alternative ways
\begin{align*}
\text{A}:\quad y_{ij}&\sim N(\zeta_j,\sigma^2), \quad \zeta_j\sim N(\alpha, \tau^2) \\
\text{B}:\quad y_{ij}&\sim N(\alpha+\zeta_j,\sigma^2), \quad \zeta_j\sim N(0, \tau^2)
\end{align*}
Version A corresponds to the two-stage formulation of hierarchical models \cite{raudenbush:2002}, where  latent variables (or random effects) appear in the model for $y_{ij}$
and  become ``outcomes'' in a second-stage model. In SEM, the first-stage and second-stage models are referred
to as the measurement part and structural part, respectively. In the Bayesian literature, this formulation has been referred to as hierarchical centering \cite{gelfand1995}. If the model is formulated like this, the hyperparameters $\bm{\psi}$ include the mean $\alpha$ and variance $\tau^2$.
Version B above specifies the reduced
form model for $y_{ij}$ so that $\zeta_j$ becomes a disturbance with zero mean and there is now only one hyperparameter, $\psi=\tau^2$.

\section{Conditional and marginal information criteria}
To illustrate problems and differences associated with the use of conditional (vs marginal) likelihoods, we focus on two specific metrics: DIC and WAIC. The former is included due to its popularity (related to its inclusion in BUGS and JAGS), whereas the latter is included because it is a more recent metric that is closer to the current ``state of the art'' in Bayesian model assessment \cite<e.g.,>{gelcar13,gelhwa14,mce15,vehtari2017practical}. 

\subsection{DIC}

The deviance information criterion (DIC) was introduced by \citeA{spibes02}.  For
a model
with parameters estimated as $\tilde{\bm{\theta}}$, they define the loss
in assigning the model-implied {\em predictive} density $f(\bm{y}^\R|\tilde{\bm{\theta}})$ to new, replicate
(or out-of-sample) data $\bm{y}^\R$  as $-2\log f(\bm{y}^\R|\tilde{\bm{\theta}})$.
Following Efron (1986), they express the expectation of this out-of-sample loss over the
distribution of the replicate data as the sum of the in-sample loss (evaluated for
the observed data) and the ``optimism'' due to using the same data to estimate  $\bm{\theta}$ and evaluate the loss:
\begin{equation}
\E_{\bm{y}^\R}[-2\log f(\bm{y}^\R|\tilde{\bm{\theta}})] \ = \ -2\log f(\bm{y}|\tilde{\bm{\theta}}) + \text{optimism},
\label{eq:explos}
\end{equation}
where the first term is sometimes referred to as the ``plug-in deviance.''
 \citeA{spibes02} use heuristic arguments to approximate the optimism by
  $2p_\D$, where
 \begin{equation}
 p_\D\ =\ \E_{\bm{\theta}|\bm{y}}[-2\log f(\bm{y}|{\bm{\theta}})] + 2\log f(\bm{y}|\tilde{\bm{\theta}}),
 \label{eq:pd}
 \end{equation}
the posterior expectation of the deviance minus the deviance evaluated at $\tilde{\bm{\theta}}$, typically the posterior expectation of the parameters. Here $p_\D$ can be interpreted as a measure of the effective number of parameters.
Combining these expressions, the DIC becomes
\begin{align}
\text{DIC}&\ = \ -2\log f(\bm{y}|\tilde{\bm{\theta}}) + 2p_\D \nonumber \\
& \ = \E_{\bm{\theta}|\bm{y}}[-2\log f(\bm{y}|{\bm{\theta}})] + p_\D
\label{eq:DIC}
\end{align}

\citeA{plummer2008penalized} proposes an alternative approximation for the optimism, which makes use of the undirected Kullback-Leibler divergence \cite<see also Plummer's contribution to the discussion of>{spibes02}. In the general case, this alternative is given by
\begin{equation}
\label{eq:plumopt}
2p_\D^P = \E_{\bm{\theta}|\bm{y}} \left [ \log \left \{ \frac{f(\bm{y}^{\R1} | \bm{\theta}^1)}{f(\bm{y}^{\R1} | \bm{\theta}^2)} \right \} + \log \left \{ \frac{f(\bm{y}^{\R2} | \bm{\theta}^2)}{f(\bm{y}^{\R2} | \bm{\theta}^1)} \right \} \right ],
\end{equation}
where $\bm{\theta}^1$ and $\bm{\theta}^2$ are independent posterior draws (e.g., from parallel chains), $\bm{y}^{\R1}$ is a replicate drawn from $f(\bm{y} | \bm{\theta}^1)$, and $\bm{y}^{\R2}$ is defined similarly. For exponential family models, including the multivariate normal that is used later, the Kullback-Leibler divergence has a closed form so that the posterior expectation can be evaluated directly without the need for replicates. However, instead of substituting the above expression for the optimism in~\eqref{eq:explos}, or in other words for $2p_D$ in the first line of~\eqref{eq:DIC} as apparently recommended by \citeA{plummer2008penalized} and as implemented in blavaan (the approach that will be called the ``Plummer definition'' in Section 4), JAGS substitutes half this expression for $p_D$ in the second line of~\eqref{eq:DIC}. The JAGS approach is convenient
as it does not require the postprocessing step of computing the plug-in deviance after the MCMC samples have been obtained. However, the JAGS approach to DIC in~\eqref{eq:plumopt} differs from the BUGS approach in~\eqref{eq:pd}, and many researchers fail to realize this difference.

\citeA{spibes02} emphasize that, for hierarchical Bayesian models, the definition of $p_\D$ and DIC depends
on which parameters are ``in focus,'' or in other words, which parameters are included in $\bm{\theta}$.  If the parameters in focus include the latent variables $\bm{\zeta}$, then $f(\cdot|\cdot)$ in Equations (\ref{eq:explos}) to (\ref{eq:DIC})
is  $f_\tc(\bm{y}| \bm{\omega}, \bm{\zeta})$ from (\ref{eq:likec}), whereas  it is $f_\tm(\bm{y}| \bm{\omega}, \bm{\psi})$ from  (\ref{eq:likem})  if the parameters in focus exclude the latent variables.
We therefore define the conditional DIC as
\begin{align}
\text{DIC}_\tc &\ = \ -2\log f_\tc(\bm{y}|\tilde{\bm{\omega}},\tilde{\bm{\zeta}}) + 2p_{\D\tc} \nonumber \\
& \ = \E_{\bm{\omega},\bm{\zeta}|\bm{y}}[-2\log f_\tc(\bm{y}|\bm{\omega},\bm{\zeta})] + p_{\D\tc}
\label{eq:DICc}
\end{align}
and the marginal DIC as
\begin{align}
\text{DIC}_\tm &\ = \ -2\log f_\tm(\bm{y}|\tilde{\bm{\omega}},\tilde{\bm{\psi}}) + 2p_{\D\tm} \nonumber \\
& \ = \E_{\bm{\omega},\bm{\psi} |\bm{y}}[-2\log f_m(\bm{y}|\bm{\omega},\bm{\psi})] + p_{\D\tm}
\label{eq:DICm}
\end{align}
with corresponding conditional and marginal variants of (\ref{eq:pd}) or (\ref{eq:plumopt}) for $p_{\D\tc}$ and $p_{\D\tm}$.
Note that Celeux et al.'s (2006)\nocite{celeux2006deviance} %
 $\text{DIC}_7$ and $\text{DIC}_1$ correspond to our conditional and marginal DIC, respectively.

Conditioning on  $\bm{\zeta}$ implies that the new units $\bm{y}^\R$ in hypothetical replications  come from the same
clusters as the observed data, whereas marginalizing over $\bm{\zeta}$ implies that the new units are for new
clusters. We show in Appendix A that
$\E_{\bm{\omega},\bm{\zeta}|\bm{y}}[-2\log f_\tc(\bm{y}|\bm{\omega},\bm{\zeta})] \leq \E_{\bm{\omega},\bm{\psi} |\bm{y}}[-2\log f_\tm(\bm{y}|\bm{\omega},\bm{\psi})]$
and in Appendix B that
$p_{\D\tc}$ tends to be much larger than $p_{\D\tm}$.
Because the magnitudes of these differences will vary between models, there will be many instances where conditional DICs and marginal DICs favor different models.
\citeA{plummer2008penalized} shows that the large-sample behavior of the penalty
$p_\D$ depends on the dimension of $\bm{\theta}$ and that the penalty may not approach the optimism
if the dimensionality of $\bm{\theta}$ increases with the sample size, as it does for latent
variable models when $\bm{\zeta}$ is in focus. This finding implies that the penalty term is a poor approximation
of the optimism when the conditional DIC is used.

When the individual clusters are not of intrinsic interest or when we would like to choose
among models that differ in the specification of the  distribution for the latent variables $p(\bm{\zeta}|\bm{\psi})$, then clearly the marginal DIC should be used.
However, the conditional DIC is easier to compute because it does not require integration. For this reason,
 users of MCMC software (e.g., BUGS, JAGS) usually obtain the conditional DIC as output
and are often unaware of the important distinction between conditional and marginal versions of the DIC.

It is sometimes possible to evaluate the marginal likelihoods by exploiting software for maximum likelihood estimation. For example, blavaan uses the related lavaan package \cite{lavaan} to evaluate the marginal likelihood  for each MCMC iteration to obtain the posterior expectation of the deviance for $p_{D_m}$ and once after MCMC sampling is complete to compute the plug-in deviance.
In the case of normal likelihoods, as in linear mixed models or SEMs with continuous responses,
the marginal likelihood has a closed form. However, closed-form solutions are generally not available for non-normal likelihoods, as in models with categorical responses.  For such cases, we propose approximating the integrals using a computationally efficient form of adaptive quadrature \cite{naylor1982applications,rabe2005maximum}. With posterior samples of the model parameters and latent variables as input, the marginal deviances
evaluated at each draw of the model parameters and at their posterior means are obtained by using the {\em marginal} (with respect to the model parameters) posterior first and second moments of the latent variables in the adaptive quadrature approximation. This quadrature method is less computationally-intensive than typical adaptive quadrature methods, because it takes place after MCMC sampling and can use all posterior samples of model parameters for the ``adaptive'' step. While the method will eventually become computationally prohibitive as the number of latent variables increases, its performance is improved over traditional quadrature methods. See Appendix C for details, where we also point out that the normal approximation to
the marginal posterior distribution could be used as the importance distribution in Monte-Carlo integration, which would be an improvement over the standard Monte-Carlo integration used by \citeA{li2016approximating} to compute predictive information criteria. For probit models and MCMC sampling with augmented variables, \citeA<>[p. 190]{fox2010} and \citeA{zhang2019} exploit the closed form of the marginal likelihood of the augmented variables to obtain different kinds of marginal DICs.

\citeA{zhaosev2017} relatedly describe and compare a variety of other methods for computing integrated likelihoods. Our approach is closest to their ``direct method'' of importance sampling (see their Section 3.1.2), with importance density chosen to be similar to what they call the ``weighted likelihood function.'' However, in the weighted likelihood function, we use the marginal prior instead of the conditional prior distribution to improve computational efficiency. In addition, we use Gauss-Hermite quadrature instead of Monte Carlo integration because the importance distribution is normal. For this reason, the \citeA{naylor1982applications} approach is closest to ours in the Bayesian literature.


\subsection{WAIC}

\citeA{gelhwa14} and \citeA{vehtari2017practical} describe  WAIC  \cite{wat10}
as an approximation to  minus twice the expected log pointwise predictive density (elppd) for new data. The definition of the WAIC
requires that the data points, such as responses $y_{ij}$ for units, are independent
given the parameters, so we first consider the conditional case where  the parameters $\bm{\theta}$ include the latent variables. In this case, minus twice elppd is
\begin{equation}
-2\sum_{j=1}^J\sum_{i=1}^{n_j}\E_{y_{ij}^\R} \log \left[\E_{\bm{\omega},\bm{\zeta}|\bm{y}} f_\tc({y}_{ij}^\R|\bm{\omega},\bm{\zeta}_j)\right] = -2\sum_{j=1}^J\sum_{i=1}^{n_j} \log \left[\E_{\bm{\omega},\bm{\zeta}|\bm{y}} f_\tc({y}_{ij}|\bm{\omega},\bm{\zeta}_j)\right] + \text{optimism}.
\label{eq:elpdd}
\end{equation}
where $\E_{\bm{\omega},\bm{\zeta}|\bm{y}} f_\tc({y}_{ij}^\R|\bm{\omega},\bm{\zeta}_j)$ on the left is the pointwise
(posterior) predictive density for a \textit{new} response $y_{ij}^r$ and its logarithm is averaged over the distribution of new responses. In contrast,
the first term on the right is evaluated at the \textit{observed} response $y_{ij}$ and therefore represents the
in-sample version of the elppd,
called lppd by \citeA{gelhwa14}. Unlike the plug-in deviance in the DIC$_\tc$, where
we condition on point estimates $\tilde{\bm{\omega}},\tilde{\bm{\zeta}}$, here we take the expectation of  $f_\tc({y}_{ij}^\R|\bm{\omega},\bm{\zeta}_j)$
over the posterior distribution of $\bm{\omega}$ and
$\bm{\zeta}$. This can be written as
\[  \E_{\bm{\omega},\bm{\zeta}|\bm{y}} f_\tc({y}^\R_{ij}|\bm{\omega},\bm{\zeta}_j) = \int f_\tc({y}^\R_{ij}|\bm{\omega},\bm{\zeta}_j) \left[ \int p(\bm{\zeta}_j | \bm{y}_j, \bm{\omega}, \bm{\psi}) p(\bm{\omega},\bm{\psi}|\bm{y}) \di\bm{\psi} \right] \di\bm{\omega}\di\bm{\zeta}_j,\]
where the term in square brackets evaluates to the joint posterior $p(\bm{\omega},\bm{\zeta}_j|\bm{y})$.
It is clear from this expression that, even after conditioning on $\bm{\omega}$  and $\bm{\psi}$, whose posterior depends on all the data,  the data  $\bm{y}_j$ for cluster $j$ provides direct information on $\bm{\zeta}_j$.
 As discussed by \citeA{gelman1996}, this \textit{posterior predictive density} is appropriate for the situation where
 hypothetical new data are responses from the existing clusters.
 The conditional WAIC is then given by
 \begin{equation}
 \text{WAIC}_\tc = -2\sum_{j=1}^J\sum_{i=1}^{n_j} \log \left[\E_{\bm{\omega},\bm{\zeta}|\bm{y}} f_\tc({y}_{ij}|\bm{\omega},\bm{\zeta}_j)\right] + 2p_{\W\tc},
 \end{equation}
 where  $p_{\W\tc}$ is the effective number of parameters and can be approximated by (see \citeA{gelhwa14} for an alternative approximation)
 \[
 p_{\W\tc} \ = \ \sum_{j=1}^J\sum_{i=1}^{n_j}  \var_{\bm{\omega},\bm{\zeta}|\bm{y}} \left[ \log f_\tc({y}_{ij}|\bm{\omega},\bm{\zeta}_j)\right].
 \]
 \citeA{vehtari2017practical} discuss problems with this approximation when any of the posterior variances of the log-pointwise predictive densities,
 $\var_{\bm{\omega},\bm{\zeta}|\bm{y}} [\log f_\tc({y}_{ij}|\bm{\omega},\bm{\zeta}_j)]$, are too large, giving 0.4 as an upper limit based on simulation evidence.

 If the hypothetical new data come from new clusters with new values ${\bm{\zeta}}_j^\R$ of the latent variables, the \textit{mixed predictive density} \cite{gelman1996,marshall2007identifying} should be  used,
  which is the  posterior expectation of $f_\tm({y}^\R_{ij}|\bm{\omega},\bm{\psi})$  and can be written as
\begin{equation}  \E_{\bm{\omega},\bm{\psi}|\bm{y}} f_\tm({y}^\R_{ij}|\bm{\omega},\bm{\psi}) = \int \left[ \int f_\tc({y}^\R_{ij}|\bm{\omega},\bm{\zeta}_j) p(\bm{\zeta}_j | \bm{\psi}) \di\bm{\zeta}_j \right] p(\bm{\omega},\bm{\psi}|\bm{y}) \di\bm{\omega}\di\bm{\psi},
\label{eq:mixedpp}\end{equation}
where the term in brackets, which evaluates to $f_\tm({y}^\R_{ij}|\bm{\omega},\bm{\psi})$, involves only the prior of $\bm{\zeta}_j$. When making predictions for a new cluster,
the data provide information on $\bm{\zeta}_j$ only indirectly by providing information on $\bm{\psi}$.
 For clustered data, we cannot simply use~\eqref{eq:mixedpp} to define the marginal WAIC because the responses
 for different units from the same cluster
 are not conditionally independent given $\bm{\omega}$ and $\bm{\psi}$, so that we must redefine the ``data points'' as
 the vectors of responses $\bm{y}_j$ for each cluster $j$. The marginal WAIC therefore becomes
 \[
  \text{WAIC}_\tm \ = \ -2\sum_{j=1}^J \log \left[\E_{\bm{\omega},\bm{\psi}|\bm{y}} f_\tm(\bm{y}_{j}|\bm{\omega},\bm{\psi})\right]  + 2p_{\W\tm}
 \]
 with
 \[
  p_{\W\tm} \ = \ \sum_{j=1}^J \var_{\bm{\bm{\omega},\bm{\psi}}|\bm{y}}  \left[ \log f_\tm(\bm{y}_{j}|\bm{\omega},\bm{\psi})\right],
 \]
 where $f_\tm(\bm{y}_{j}|\bm{\omega},\bm{\psi})$ is given as term $j$ of the product in~\eqref{eq:likem}.
 \citeA{li2016approximating} and \citeA{millar2018} define WAIC$_\tm$ for the case when there is only one unit per cluster. Examples are overdispersed Poisson or binomial regression models, where
 the unit-level latent variable modifies the variance function, and meta-analysis, where the data for the clusters have been aggregated into effect-size estimates. 
 
\subsection{Interpretation of WAIC via connections with LOO-CV}
\citeA{wat10} showed that WAIC and  Bayesian leave-one out (LOO) cross-validation (CV) are asymptotically equivalent. Following our
previous discussion, the different forms of WAIC correspond to different forms of LOO-CV.
Because the predictive distribution in the conditional WAIC is for a response from a new unit
in an existing cluster, it approximates leave one \textit{unit} out CV, given as
\[ -2 \text{LOuO-CV} = -2 \displaystyle \sum_{j=1}^J \displaystyle \sum_{i=1}^{n_j} \log \E_{\bm{\omega},\bm{\zeta}|\bm{y}_{-ij}} f_\tc({y}_{ij}|\bm{\omega},\bm{\zeta}_j), \]
where $\bm{y}_{-ij}$ represents the full data with observation $ij$ held out.
In contrast, the predictive distribution in the marginal WAIC is
the joint marginal distribution for the responses of all units in a new cluster and therefore approximates
 leave one \textit{cluster} out CV:
\[ -2 \text{LOcO-CV} = -2 \displaystyle \sum_{j=1}^J \log \E_{\bm{\omega},\bm{\psi}|\bm{y}_{-j}} f_\tm(\bm{y}_{j}|\bm{\omega},\bm{\psi}). \]
 In other words, the marginal WAIC generalizes to new clusters
whereas the conditional WAIC generalizes only to new units in the clusters that are in the data.
Note that the first term in WAIC$_\tm$, the in-sample lppd, corresponds to  the ``hierarchical approximation'' \cite{vehtari2016} 
or the ``full-data mixed''  \cite{marshall2007identifying} method for approximating what we call LOcO-CV.

\citeA{li2016approximating} and \citeA{millar2018} motivate their marginal WAIC$_\tm$ as an approximation to LOO-CV in the case
of one unit per cluster.  \citeA{millar2018} shows that conditional leave-one out (our LOuO) becomes marginal leave-one out (our LOcO) when there is only one unit per cluster,
because there are no data for the cluster to condition on after removing the unit. Therefore, WAIC$_\tc$ has no clear justification in this case.  \citeA{vehtari2017practical} nevertheless use WAIC$_\tc$ for the ``eight schools'' data, a meta-analysis
of SAT preparatory programs. They find that WAIC$_\tc$ diverges from LOO-CV when the responses are multiplied by a factor $S$ with $S>1.5$ (see their Figure 1a).
We replicated their results and found that WAIC$_\tm$, which \citeA{vehtari2017practical} did not consider, continues to be a good approximation to LOO-CV
for $S=4$, for which we obtained LOO-CV$=86.0$, WAIC$_\tm=85.5$, and WAIC$_\tc=68.7$. \citeA{millar2018} also points out that two regularity conditions required for asymptotic equivalence
of WAIC and LOO-CV do not hold for the conditional versions, namely (1) the observations are not identically distributed (their distribution depends on the latent variables $\bm{\zeta}_j$) and (2)
the number of parameters increases with the sample size. These two regularity conditions are also violated in the clustered case, so it is not clear whether WAIC$_\tc$ is a good approximation to LOuO-CV.

Recent work describes the fact that LOO-CV (and, consequently, WAIC) does not necessarily select the true or best model. \citeA{piiveh2017} explored use of WAIC (and other metrics) for selecting subsets of predictors in a regression context, finding that variability associated with its estimation can lead to poor model selection. \citeA{growag2018} provide three artificial examples where LOO-CV does not asymptotically select the data-generating model. Discussants of these issues have recommended against the sole use of point estimates for model selection, also considering, e.g., qualitative patterns of data that the model can accommodate \cite{nav2018} and uncertainty associated with the model's predictive accuracy \cite{vehsim2018}. Uncertainty in a model's predictive accuracy may include explicit consideration of variability in a metric like WAIC, or it may include a different procedure such as Bayesian model averaging \cite{hoemad99} or stacking \cite{yao2018}.

\citeA{vehtari2017practical} introduce a computationally efficient approximation to LOO-CV using Pareto-smoothed importance sampling (PSIS), and they implement this method in the R package loo \cite{loopackage}. This package computes both PSIS-LOO and WAIC with MCMC draws of
pointwise predictive densities as input, and it can be used to compute PSIS-LOuO, PSIS-LOcO, WAIC$_\tc$, and WAIC$_\tm$  by providing either  $\log f_\tc({y}_{ij}|\bm{\omega},\bm{\zeta}_j)$ or $\log f_\tm(\bm{y}_{j}|\bm{\omega},\bm{\psi})$ as input. This package is used in Section 4.2.


\section{Examples}
We now discuss two examples to highlight differences between various types of Bayesian information criteria. The first example involves a series of increasingly constrained multiple-group factor analysis models, as might be used in a measurement invariance study. This example focuses on DIC and its various definitions across software and across conditional/marginal likelihoods. The second example involves item response models, where marginal versions of the criteria do not have closed forms. In this example, we consider WAIC and PSIS-LOO in addition to DIC.


\subsection{Measurement Invariance Study via Multi-Group CFA}
\citeA{wicdol05} compared a series of four-group, one-factor models using data from a stereotype threat experiment. Here we replicate these model comparisons via both conditional DIC and marginal DIC; the original authors' comparisons involved frequentist models. This allows us to illustrate the practical impact of using different DICs.

\subsubsection{Method}
The data consist of Differential Aptitude Test scores from 295 Dutch students comprising both majority and minority Dutch ethnicities. The test contained three subscales (verbal ability with 16 items, mathematical ability with 14 items, and abstract reasoning with 18 items), where each subscale score was computed as the number of items that a student answered correctly. Gaussian factor analysis is used here; it would perhaps be more appropriate to employ an item factor analysis model here, but our models below generally match those of the original authors.

The original authors used a ``stereotype threat'' manipulation during the data collection: half of the students were primed about ethnic stereotypes related to intelligence tests, while the other half received no primes. \citeA{wicdol05} conducted a measurement invariance study with the resulting data, to examine the impact of the stereotype threat manipulation on the students. This measurement invariance study involved four groups: two ethnicities (majority, minority) crossed with the stereotype threat manipulation (received, not received).
The measurement model for subscale $i$ ($i=1,2,3$) and person $j$ ($j=1,\ldots, 295$) in group $g$ ($g=1,2,3,4$) is
\begin{align*}
 y_{ijg} | \nu_{ig}, \lambda_{ig}, \sigma_{ig}, \zeta_{jg} &\sim \mathrm{N}( \nu_{ig} + \lambda_{ig} \zeta_{jg}, \sigma^2_{ig}),
\end{align*}
where $\nu_{ig}$ is a group-specific intercept, $\lambda_{ig}$ is a group-specific factor loading for subscale $i$ (with the loadings for the first subscale set to 1 for identification, $\lambda_{1g}=1$), $\zeta_{jg}$ is the common factor, and $\sigma^2_{ig}$ is the group-specific variance of the unique factor for subscale $i$.

The parameters
 $\nu_{ig}$, $\lambda_{ig}$, and $\sigma_{ig}^2$ are in $\bm{\omega}$ and have the following prior distributions for
 $g=1,2,3,4$:
\begin{align*}
\nu_{1g}, \nu_{2g}, \nu_{3g} &\sim \mathrm{N}(0, 1000)\\
\lambda_{2g}, \lambda_{3g} &\sim \mathrm{N}(0, 100)\\
\sigma^{-2}_{1g}, \sigma^{-2}_{2g}, \sigma^{-2}_{3g} &\sim \mathrm{Gamma}(1, .5)
\end{align*}
The distribution of the common factor, or structural part of the SEM is
\[
\zeta_{jg} \sim \mathrm{N}(\alpha_g, \tau^2_g)
\]
with group-specific means $\alpha_g$ and variances $\tau^2_g$. These hyperparameters are in $\bm{\psi}$ and have the following priors:
\begin{align*}
\alpha_g &\sim \mathrm{N}(0, 100)\\
\tau^{-2}_g &\sim \mathrm{Gamma}(1, 1),
\end{align*}
with the exception that $\alpha_1$ is fixed to zero for identification. These prior distributions are similar to the defaults supplied by blavaan. They are just informative enough so that most models converge in JAGS, while being uninformative enough to not exert much influence on the parameters' posterior distributions. 

\begin{table}
  \caption{Description of estimated models. The parameter count is the number of free parameters in the marginal likelihood. Also see Table 3 of Wicherts et al., 2005.}
  \label{tab:wic}
  \centering
  \begin{tabular}{llc}\\\hline
    Model & Parameter Restrictions & Parameter Count \\\hline
    2 & All $\lambda_{ig} = \lambda_i\ i=1,\ldots,3; g=1,\ldots,4$ & 30 \\
    2a & Model 2, except $\lambda_{24}$ unrestricted & 31 \\
    3 & Model 2a, plus all $\sigma^2_{ig} = \sigma^2_i$ & 22 \\
    3a & Model 3, except $\sigma^2_{24}$ unrestricted & 23 \\
    4 & Model 3a, plus $\tau^2_1 = \tau^2_2$ and $\tau^2_3 = \tau^2_4$ & 21 \\
    5 & Model 4, plus all $\nu_{ig} = \nu_i$ except $\nu_{24}$ unrestricted & 16 \\
    5a & Model 5, except $\nu_{31}$ unrestricted & 17 \\
    5b & Model 5a, except $\nu_{32}$ unrestricted & 18 \\
    6 & Model 5b, plus $\alpha_1 = \alpha_3$ & 17 \\\hline
  \end{tabular}
\end{table}

The measurement invariance study involved the comparison of nine versions of the model above; specific model details are shown in Table~\ref{tab:wic}. The overall strategy was to set increasingly more parameters equal across groups (i.e., across the $g$ subscript), starting with factor loadings $\lambda_{ig}$ (models 2 and 2a), then unique factor variances $\sigma^2_{ig}$ (models 3 and 3a), then common factor variances $\tau^2_g$ (model 4;  constant across conditions but allowed to differ between minority and non-minority groups), then intercepts, $\nu_{ig}$ (models 5, 5a and 5b), and finally factor means $\alpha_g$ (model 6; constant across conditions but allowed to differ between minority and non-minority groups). The models without letters resulted from the previous model by constraining one type of parameter, whereas the model with the same number and a letter resulted from freeing some of these parameters again based on SEM fit criteria and modification indices. For example, model 5 (intercepts restricted across all four groups) is nested
 in models 5a (one intercept freed) and 5b (another intercept freed), and model 6 (factor means constrained equal across conditions) is nested in model 5b.
See \citeA{wicdol05} for further detail on the models.

We conduct a Bayesian re-analysis of the nine \citeA{wicdol05} models here, focusing on model comparison via DIC. We used JAGS to estimate each of the models ten times; for each of these estimations, we used ``automatic'' computations of chain length and convergence proposed by \citeA{raflew1995} and implemented in R package {\em runjags} \cite{den2016}. This algorithm first runs three chains for 4,000 iterations each, checking that all parameters' Gelman-Rubin statistics \cite{gelrub92} are below 1.05. If not, the algorithm runs for 4,000 more iterations and re-checks, continuing in this fashion until either the statistics are below 1.05 or a maximum time limit has been reached. Once the 1.05 threshold is achieved for all parameters, the algorithm performs a computation to determine the number of additional samples necessary to estimate each parameter's 2.5th quantile to an accuracy of .005 with probability .95. Then the chains are run for this number of additional samples. Thus, the specific number of samples varies for each model estimation.

For each estimation, we computed four DIC statistics via blavaan: the conditional and marginal versions using both the  \citeA{spibes02} definition in~(\ref{eq:pd}) and the \citeA{plummer2008penalized} definition in~(\ref{eq:plumopt}) for ``effective number of parameters.'' 


\subsubsection{Results}
\begin{figure}
\caption{Marginal and conditional DICs (Spiegelhalter et al.\ definitions) for nine models from Wicherts et al., where each model was estimated ten times.}
\label{fig:wicdic}
\begin{center}
\includegraphics[width=5in]{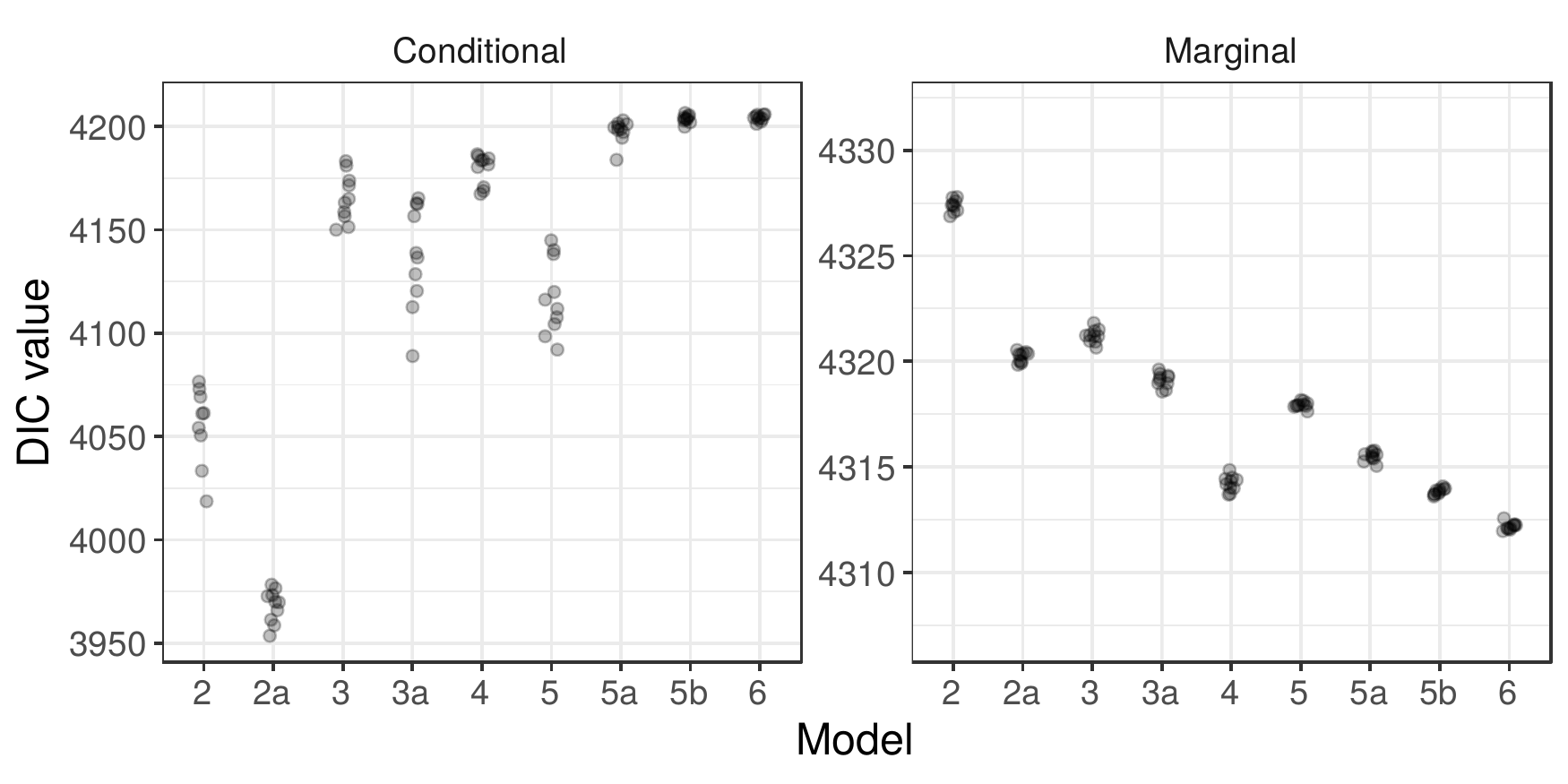}
\end{center}
\end{figure}

Spiegelhalter et al.'s (2002) conditional and marginal DICs for ten estimations of each of the nine models are displayed in Figure~\ref{fig:wicdic}.
Here the scale of the $y$ axis for the marginal case is a 10-fold magnification of that for the conditional case, showing that the Monte Carlo error is far greater for
the conditional DICs. However, there is substantial Monte Carlo error also in the marginal case, making model comparison unreliable for some pairs of models (e.g., Models 3 and 2a). This suggests that accurate DICs require sampling for more iterations, as compared to what is required for obtaining accurate parameter estimates.

Especially because these models were part of a measurement invariance study, we also examine where we would stop as we moved from Model 2 to 6 in a sequential/stepwise fashion (as was done in the original paper, though this is not required). In the marginal case, the DIC for Model 3 is generally larger than the DIC for Model 2a, so we may stop at Model 2a. However, Monte Carlo error is large enough that we may prefer Model 3 to Model 2a in a given replication, in which case we may stop at Model 4. But the marginal DIC is lowest for the final Model 6. The results are very different in the conditional case. Here, we would stop at Model 2a across all ten replications. Further, the models generally obtain larger  DIC values as we move towards the final model 6, in stark contrast to the marginal case. Models 5b and 6 have the worst values in the conditional case, whereas they are the best in the marginal case.

\begin{figure}
\caption{Estimated effective number of parameters under marginal and conditional DIC for nine models from Wicherts et al., where each model was estimated ten times. Lines reflect simple parameter counts for each model, where latent variables count as parameters in the conditional case. The Plummer DIC computes $p_\D$ via Kullback-Leibler divergence as in~\eqref{eq:plumopt}, whereas the Spiegelhalter DIC computes $p_\D$ via deviances as in~\eqref{eq:pd}.}
\label{fig:wicpd}
\begin{center}
\includegraphics[width=5in]{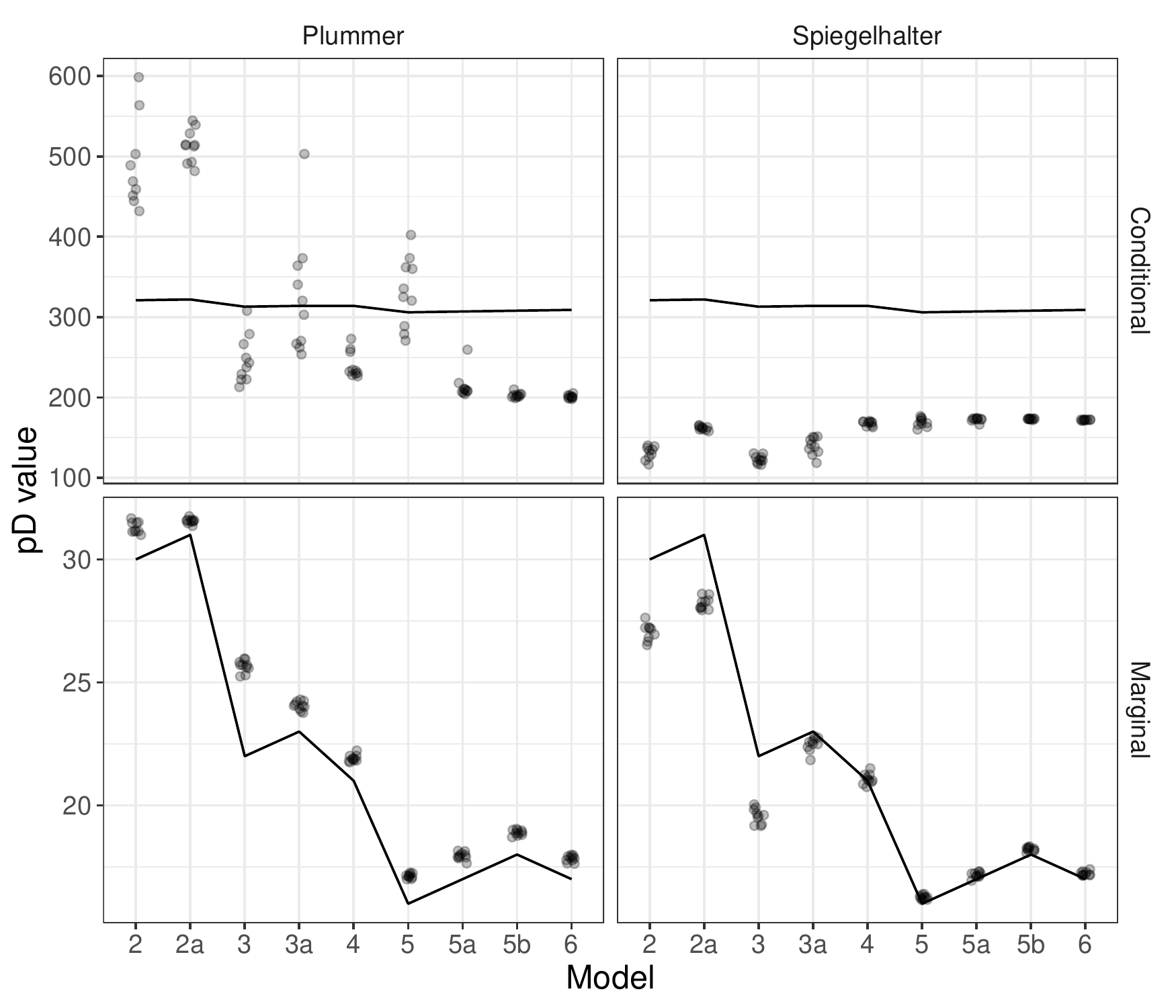}
\end{center}
\end{figure}

Beyond the comparison of DICs, we further examine the four definitions of effective number of parameters: marginal versus conditional crossed with the Spiegelhalter versus Plummer definitions. These are displayed for the \citeA{wicdol05} models in Figure~\ref{fig:wicpd}, with the lines showing simple parameter counts for each model (where, in the conditional case, latent variable values for the students count as parameters). Results are similar to the previous figure in that the conditional definitions exhibit greater Monte Carlo error than the marginal definitions (note the differences in $y$-axis scales, both within this figure and as compared to Figure~\ref{fig:wicdic}). In the marginal case, we are counting the unrestricted item-specific parameters in $\bm{\omega}$ and the unrestricted factor means and variances in $\bm{\psi}$  (note that these parameter counts increase when parameters are freed, i.e. going from model 2 to 2a, 3 to 3a, and from 5 to 5a to 5b.); the Plummer effective number of parameters is generally larger than the Spiegelhalter effective number of parameters here. In the conditional case,
we are counting the unrestricted item-specific parameters in $\bm{\omega}$ and the common factor values for all students. Here we see greater discrepancy between the Spiegelhalter and Plummer definitions. The Spiegelhalter definition appears to work better here, as the values are nearly always smaller than the associated parameter counts (which we would expect based on the fact that the latent variables exhibit shrinkage). However, in fairness, \citeA{plummer2008penalized} explicitly states that his definition is not intended for situations where the number of parameters increases with the number of observations, as they do in the conditional case. Beyond this, we observe situations where the marginal effective number of parameters decreases while the conditional effective number of parameters increases or stays the same; the sequence from Models 4 to 6 exhibits multiple examples.

\begin{figure}
\caption{Marginal and conditional DICs (Plummer definitions) for ten replications of the Wicherts et al.\ models, with error bars ($\pm$ 2 SDs) stemming from a single replication.}
\label{fig:jagerr}
\begin{center}
\includegraphics[width=5in]{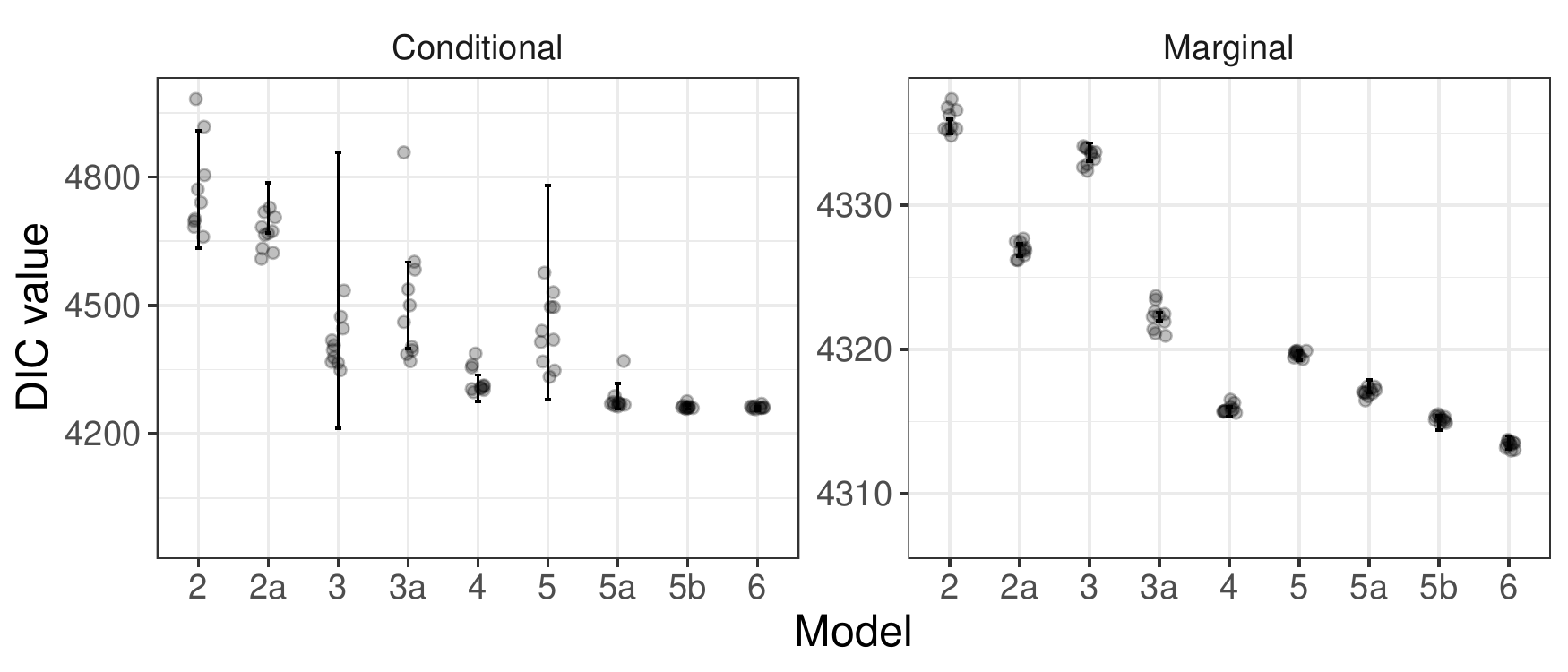}
\end{center}
\end{figure}

Finally, we point out that much of the variability in DIC is due to Monte Carlo error in the estimate of the effective number of parameters. As shown in Appendix D, this Monte Carlo error is easy to estimate (in the Plummer case by ignoring any Monte Carlo error in the plug-in deviance). Figure~\ref{fig:jagerr} plots the Plummer DICs across the 10 replications, with error bars representing $\pm 2$ times the estimated Monte Carlo error (from a single replication) only in the effective number of parameters. We observe that the error bars reflect the magnitude of variability in the overall DIC values. To support this claim,  we computed the proportion of the time that the error bars  covered the DICs across the ten replications and nine models. This coverage was 76\% in the conditional case and 83\% in the marginal case (the coverage for the corresponding effective numbers of parameters was 89\% and 99\%, respectively).  While the coverage for the DIC is lower than the nominal 95\%, it illustrates that the deviance evaluated at the posterior means exhibits small variability compared to the variability in expected deviance (Spiegelhalter case) or K-L distance (Plummer case).

\subsubsection{Prior Sensitivity}
While the prior distributions in the previous section were related to blavaan default settings, other sets of priors could be considered. Two plausible sets include (i) less-informative priors, closer to {\em Mplus} defaults, and (ii) informative priors based on substantive knowledge of the data. We conducted the same analyses described in the previous section under these two new sets of priors. 

The less-informative set of priors was
\begin{align*}
\nu_{1g}, \nu_{2g}, \nu_{3g} &\sim \mathrm{N}(0, 10,000)\\
\lambda_{2g}, \lambda_{3g} &\sim \mathrm{N}(0, 1000)\\
\sigma^{-2}_{1g}, \sigma^{-2}_{2g}, \sigma^{-2}_{3g} &\sim \mathrm{Gamma}(.01, .01)\\
\alpha_g &\sim \mathrm{N}(0, 1000)\\
\tau^{-2}_g &\sim \mathrm{Gamma}(.1, .1)
\end{align*}
while the informative set of priors was
\begin{align*}
\nu_{1g}, \nu_{2g}, \nu_{3g} &\sim \mathrm{N}(7, 10)\\
\lambda_{2g}, \lambda_{3g} &\sim \mathrm{N}(0, 1)\\
\sigma^{-2}_{1g}, \sigma^{-2}_{2g}, \sigma^{-2}_{3g} &\sim \mathrm{Gamma}(2.5, 5)\\
\alpha_g &\sim \mathrm{N}(0, 1)\\
\tau^{-2}_g &\sim \mathrm{Gamma}(2.5, 5).
\end{align*}
The informative priors were selected so that the posterior density was generally in the range of plausible parameter values: intercept priors are chosen to reflect the fact that the observed variables are scale scores that range between 0 and 18; loading priors are based on the fact that one loading is fixed to 1, and other loadings are expected to be similar; and the gamma distributions on precisions are selected based again on the scale scores' ranges (knowing that the ranges of the scale scores cannot produce extremely large variances). 

\begin{figure}
\caption{Marginal and conditional DICs (Spiegelhalter et al.\ definitions) under informative prior distributions for nine models from Wicherts et al.}
\label{fig:wicdicstr}
\begin{center}
\includegraphics[width=5in]{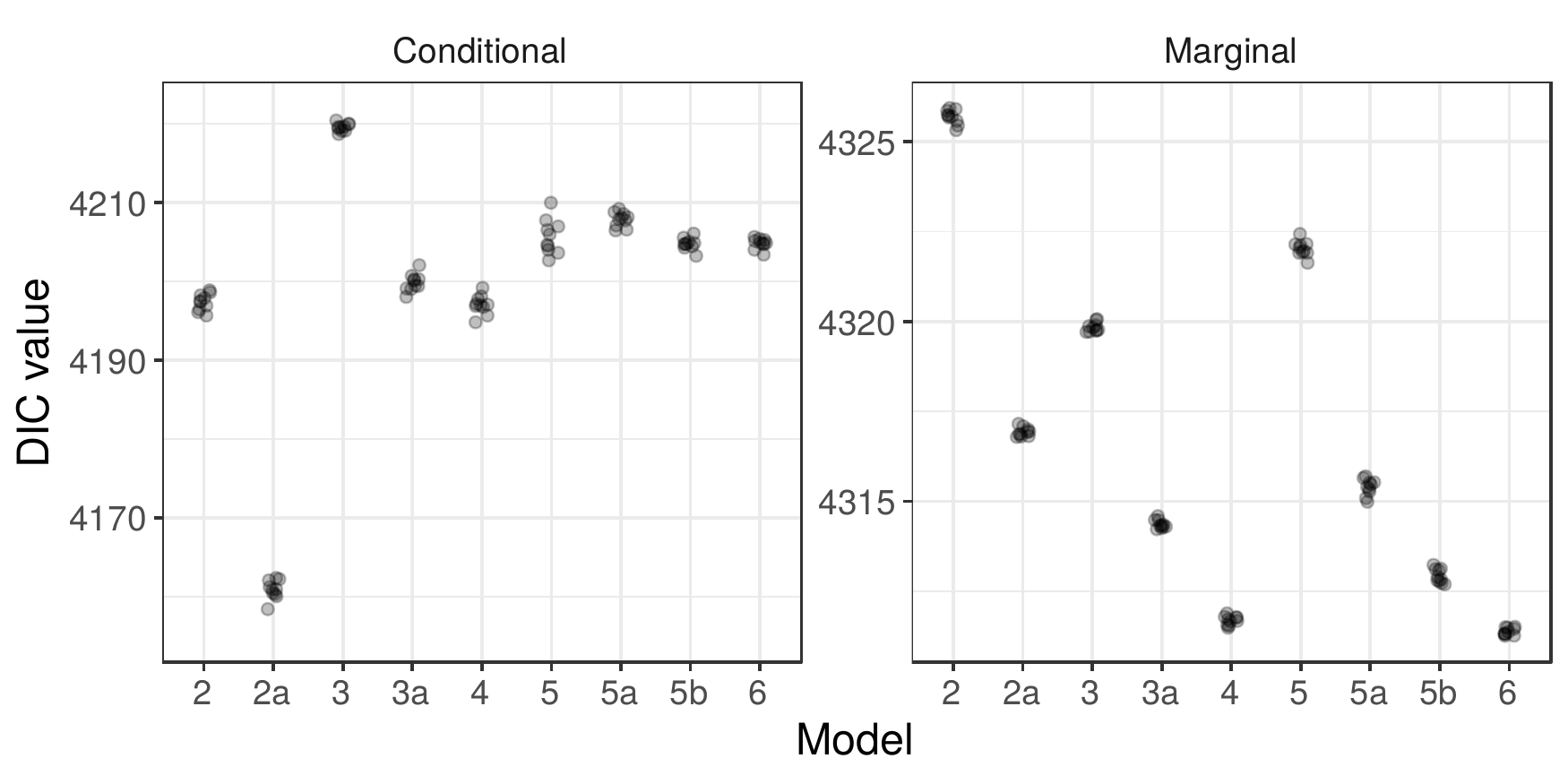}
\end{center}
\end{figure}

The results for the uninformative set of priors (shown in Appendix E) are similar to the previous results, with some additional Monte Carlo error and convergence issues. 
The results for the informative set of priors are more interesting. Figure~\ref{fig:wicdicstr} is similar to Figure~\ref{fig:wicdic}, except that the conditional y-axis scale is seven times the marginal y-axis scale (as opposed to Figure~\ref{fig:wicdic}, where marginal was ten times as large). We generally observe that the conditional metrics' Monte Carlo errors have been drastically reduced under informative priors, though the error is still larger than that of the marginal metrics. The conditional and marginal metrics' selections of the best model continue to disagree with one another. Additionally, the Plummer computation of conditional DIC prefers a different model as compared to the Spiegelhalter computation of conditional DIC (see Appendix E). 

The results in this section illustrate that the use of marginal vs conditional DIC can impact substantive conclusions, in that different models can be selected with different versions of DIC and different conclusions can be reached about the effective number of parameters. We speculate that conditional DIC will generally prefer models of greater complexity, due to the relationship between conditional/marginal DIC and cross-validation. Because conditional DIC is related to predicting a held-out unit from an existing cluster, the remaining data from that cluster are helpful for making the prediction. In contrast, when predicting data from a held-out cluster (which is related to marginal DIC), we have greater uncertainty about the specific properties of the cluster. Thus, conditional DIC may support models of greater complexity, as compared to marginal DIC.

Beyond model preference, the conditional DIC exhibits larger Monte Carlo error and depends more strongly on the approximation used (Spiegelhalter versus Plummer), as compared to the marginal version. Further, the Monte Carlo error in conditional DIC is drastically reduced through use of informative prior distributions on model parameters. While strong prior information is not always available, this result suggests that even mild prior information can aid in precise computation of DIC. In the next section, we extend these results to item response models and other information criteria.

\subsection{Explaining Individual Differences in Verbal Aggression via IRT}
Here, we consider item response models with different person covariates, comparing them via conditional and marginal versions of DIC (Spiegelhalter et al.\ definitions), WAIC, and PSIS-LOO. In the case of IRT, the marginal information criteria utilize the numerical integration methods described in Appendix C.

\subsubsection{Method}
Several latent regression Rasch models are fit to the dataset on verbal aggression \cite{Vansteelandt2000} that consists of $J = 316$ persons and $I = 24$ items.
Participants were instructed to imagine four frustrating scenarios, and for each they responded to items regarding whether they would react by cursing, scolding, and shouting. They also responded to parallel items regarding whether they would \textit{want} to engage in the three behaviors, resulting in six items per scenario (cursing/scolding/shouting $\times$ doing/wanting). An example item is, ``A bus fails to stop for me. I would want to curse.'' The response options for all items were ``yes'', ``perhaps'', and ``no.'' The items have been dichotomized for this example by combining ``yes'' and ``perhaps'' responses. Two person-specific covariates are included: the respondent's trait anger score \cite{spielberger1988state}, which is a raw score from a separate measure taking values between 11 and 39, and an indicator for whether the respondent is male, which takes the values 0 and 1. This leads us to consider five possible models: a model without person covariates, two models with one person covariate, a model with both person covariates, and a model with both person covariates plus their interaction.

The model that defines the conditional likelihood $f_c(\bm{y}|\bm{\omega},\bm{\zeta})$ is
\begin{equation} \label{eq:4-latregrasch}
  y_{ij} | \bm{x}_j, \bm{\gamma}, \zeta_j, \delta_i \sim
  \mathrm{Bernoulli}\left (
    \mathrm{logit}^{-1}(\bm{x}_j' \bm{\gamma} + \zeta_j - \delta_i)
  \right ),
\end{equation}
where $y_{ij} = 1$ if person $j$ responds positively to item $i$  and $y_{ij} = 0$ otherwise, $\bm{x}_j$ is a vector of a constant
and $K-1$ person-specific covariates, $\bm{\gamma}$ is a vector of regression coefficients, $\zeta_j$ is a person residual, and $\delta_i$ is an item difficulty parameter. The last item difficulty is constrained such that $\delta_I = -\sum_{i}^{(I-1)} \delta_i$. The priors for the item difficulties and regression coefficients that comprise $\bm{\omega}$
are
\begin{align*}
  \delta_1, \ldots, \delta_{I-1} &\sim \mathrm{N}(0, 9)\\
   \gamma_1^*,\ldots,\gamma_K^* &\sim t_1(0, 1).
\end{align*}
Unlike the previous example, $\zeta_j$ is just a disturbance here with zero mean and hyperparameter $\psi=\tau$:
\begin{align*}
 \zeta_j & \sim \mathrm{N}(0, \tau^2)\\
  \tau & \sim \mathrm{Exp}(.1).
\end{align*}

Weakly informative priors are specified for the regression coefficients $\bm{\gamma}$, following \citeA{gelman2008weakly}.
Specifically, after standardizing continuous covariates to have zero mean and standard deviation 0.5 and binary covariates
to have mean 0 and range 1, $t$-distributions with one degree of freedom are used as priors for the corresponding regression coefficients,
$\bm{\gamma}^*$. The MCMC draws of the coefficients are then transformed to reflect the original scales of the covariates. Next, the priors for item difficulties have about 95\% of their density lying between $(-6,6)$, reflecting the fact that the difficulties are on the logit scale (so it would be surprising to observe values outside of this range). Finally, the prior on $\tau$ is simply chosen to be uninformative.

Focus is placed on $\zeta_j$ for the conditional approach, which yields a prediction inference involving new responses from the same persons (and items). The marginal approach, perhaps more realistically, places focus on $\tau$, implying a prediction inference involving new responses from new people.
Five competing models are considered, differing only in what person covariates are included: Model 1 includes no covariates ($K=1$), Model 2 has the trait anger score ($K=2$), Model 3 has the indicator for male ($K=2$), Model 4 has both covariates ($K=3$), and Model 5 has both covariates and their interaction ($K=4$).

\subsubsection{Results}

The five models are estimated via Stan using 5 chains of 2,500 draws with the first 500 draws of each discarded, resulting in a total of
10,000
kept posterior draws. The unusually large number of posterior draws (for Hamiltonian Monte Carlo) is chosen here due to the anticipated Monte Carlo errors for the information criteria, but such a large number is not ordinarily necessary for estimating the posterior means and standard deviations of the parameters.
The adaptive quadrature method described in Appendix C is used to integrate out $\bm{\zeta}$ for computing the marginal versions of DIC, WAIC, and PSIS-LOO.
Eleven quadrature points were found to be sufficient for one of the models (Model~4) using the method described in Appendix C, and this number of points is used throughout. On a Windows desktop with 4 cores and 16GB of memory, the quadrature method took about 90 seconds per model to complete (with parallel processing).

\begin{figure}
\caption[Information criteria for the five latent regression Rasch models]
{Information criteria for the five latent regression Rasch models. Points represent the results of the $10$ independent MCMC simulations per model. A small amount of horizontal jitter is added to the points. }
\label{fig:3-example-ic}
\begin{center}
\includegraphics{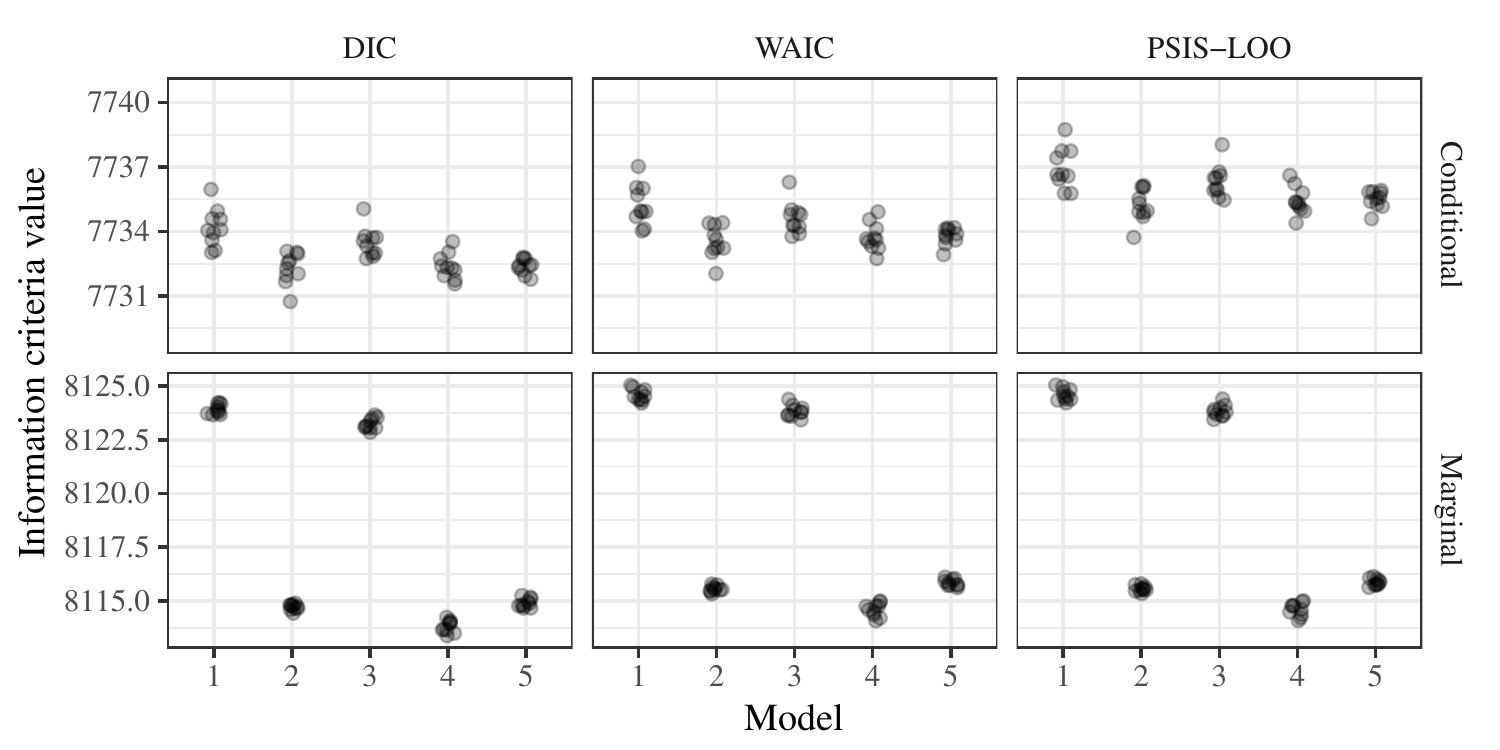}
\end{center}
\end{figure}

Figure~\ref{fig:3-example-ic} provides the estimates for the information criteria and PSIS-LOO, where the conditional and marginal versions now have the same $y$-axis scales.  The DIC and WAIC differ from each other for any given model, though they seem to show a similar pattern between models. The WAIC is a much better approximation to the PSIS-LOO in the marginal case than the conditional case. The high degree of Monte Carlo error in the conditional versions renders differentiating the predictive performance of the models difficult. In the marginal case, the amount of Monte Carlo error is less but still poses a degree of difficulty in making comparisons. However, Models~1 and 3 now clearly provide poorer predictions in comparison to the others. These models do not include trait anger, suggesting that this variable is an important predictor of verbal aggression. There is some evidence supporting Model~4 (both covariates, no interaction) as the best among the candidates.

\begin{figure}[htb]
\caption[Estimated effective number of parameters for the five latent regression Rasch models]
{Estimated effective number of parameters ($\hat p$) for the five latent regression Rasch models. Points represent the results of the $10$ independent MCMC simulations per model. A small amount of horizontal jitter is added to the points. The horizontal lines represent the counts of parameters associated with each model and focus.}
\label{fig:3-effn}
\begin{center}
\includegraphics{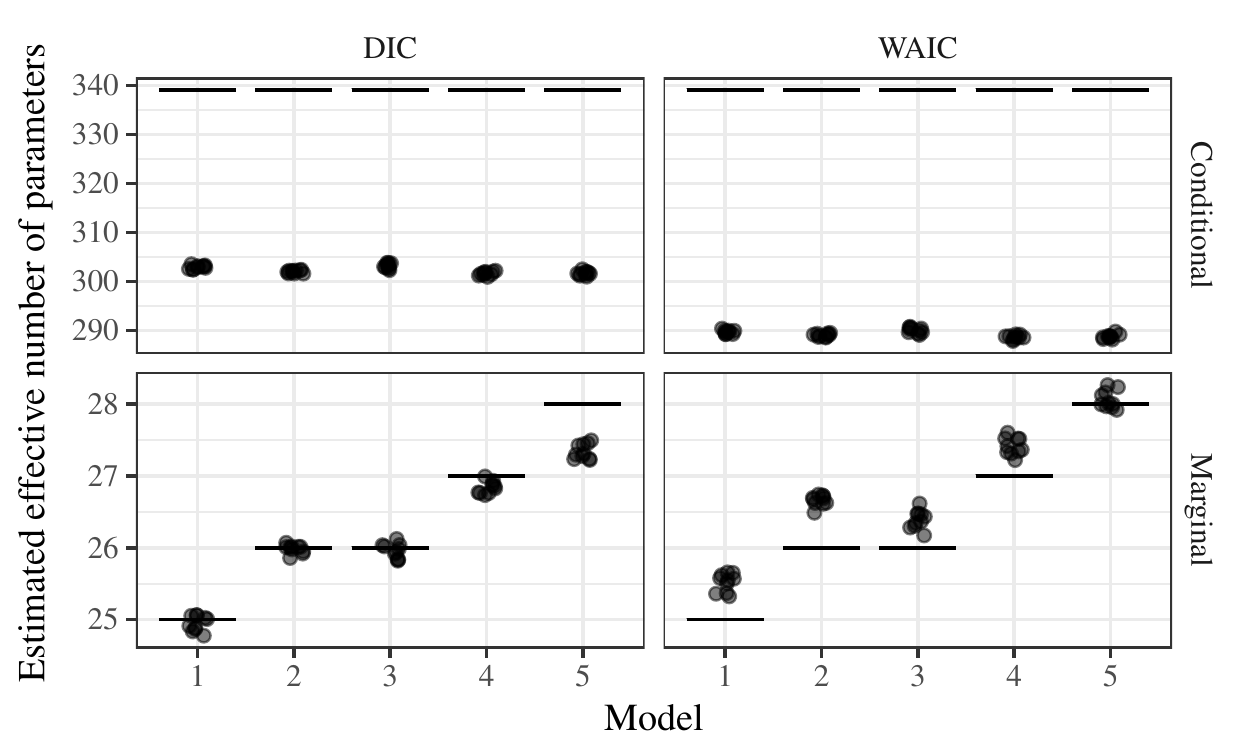} 
\end{center}
\end{figure}

Figure~\ref{fig:3-effn} provides the estimated effective number of parameters (dots) and actual number of parameters (lines) for the conditional and marginal versions of the DIC and WAIC. 
For conditional information criteria, the effective number of parameters is substantially less than the actual numbers of parameters, owing mainly to the fact that each $\zeta_j$, with its hierarchical prior, contributes less than one to the effective number of parameters. For marginal information criteria, on the other hand, the effective number of parameters are close to the actual numbers of model parameters and, for WAIC, are somewhat larger than the number of parameters.

For the WAIC, contributions to the effective number of parameters (from item-person combinations for $p_{\W\tc}$ and from persons for $p_{\W\tm}$) should be less than .4 which was always satisfied for the marginal WAIC but was violated
for an average of between 2.1 and 3.4 contributions for the conditional WAIC across the five models \cite<for details, see>{furr2017}.

\section{Discussion}
In this paper, we sought to clarify and illustrate the various forms of Bayesian information criteria that are used in practice. Researchers often rely on pre-packaged software to obtain these criteria, failing to realize the fact that multiple versions of the criteria are available depending on the software (BUGS vs.\ JAGS) and depending on whether or not the latent variables appear in the model likelihood (leading to {\em conditional} and {\em marginal} information criteria, respectively). Additionally, we showed that the criteria can suffer from large amounts of Monte Carlo error. Hence, long chains will often be necessary to obtain precise point estimates of the criteria. If ignored, these issues can lead to model comparisons that are irrelevant and/or irreproducible. While the conditional and marginal criteria will sometimes agree on the best model, the agreement is dependent on the number and types of models under consideration. In Appendices A and B, we provide theoretical results showing that the conditional and marginal information criteria will generally differ from one another.

\paragraph{Recommendations.} Similarly to the use of marginal likelihoods in general latent variable models, we recommend use of marginal information criteria as the defaults in Bayesian analyses. As previously discussed, marginal criteria assess a model's predictive ability when applied to new clusters. This is exactly what is desired in many psychometric contexts, where clusters may be defined by, e.g., countries, schools, or students. In these cases, we wish to discern general properties of scales or items that are not specific to the clusters that were observed. As a side advantage, the marginal information criteria also tend to have less Monte Carlo error than their conditional counterparts. We would expect similar issues to arise when one uses posterior predictive estimates to study model fit, and this is a topic of future study.

Because the chain length necessary to obtain precise information criteria is generally larger than the length required to obtain stable posterior means of individual parameters, it may be useful to compute the information criteria at multiple points. First, the criteria can be computed when individual parameters' posterior means have converged and stabilized. The researcher can then continue running the chains for the same number of iterations, re-computing the information criteria and effective number of parameters, and assessing changes in results. This process can be continued until the information criteria have reached the desired level of precision. Alternatively, researchers might apply the \citeA{raflew1995} automatic chain length computations directly to samples of the model deviance (these computations are typically applied to samples of individual model parameters). Regardless of strategy employed, our examples suggest that the Monte Carlo error of the effective number of parameters largely track the imprecision of the information criteria.

Beyond Monte Carlo error, there are further issues that lead us to not recommend conditional information criteria. For the DIC, \citeA{plummer2008penalized} showed that the penalty does not approach the optimism when the number
of parameters increases with the sample size (as they do in the conditional case), and we observed in the first example that the penalty term depended largely on the definition used (Plummer versus Spiegelhalter). In the second example, we observed that the WAIC differed much more from PSIS-LOO in the conditional case than in the marginal case. This corresponds to the work of \citeA{millar2018}, who points out that two regularity conditions
are not satisfied in the conditional case, one of them again being the number of parameters increasing with sample size. In the context of multilevel IRT models, \citeA{zhang2019} also recommend against the conditional version of the DIC.

\paragraph{Concluding remarks.} Researchers may also be interested in the abilities of the information criteria to select the true model. We note that the information criteria are not designed to select the true model but instead to select the model with best out-of-sample predictive accuracy. In small datasets with complex data-generating models, the model with best out-of-sample predictive accuracy will tend to be simpler than the true model, because the true model would lead to overfitting and poor out-of-sample predictive accuarcy.

\citeA{zhang2019} conducted simulation studies using DIC and obtained related results. Their studies involved multilevel IRT models where item responses were nested in individuals, and individuals were nested in schools. For these models, one can marginalize only over person parameters or over both person and school parameters, leading to different forms of marginal DIC. Additionally, the authors considered ``joint'' forms of DIC that, in our notation, employed likelihoods that are roughly of the form $f(\bm{y}_j, \bm{\zeta}_j | \bm{\omega})$. Zhang et al.\ found that the conditional DIC sometimes preferred models that were more complex than the data-generating model, and they ultimately recommended a version of DIC based on the person-level joint likelihood. This joint likelihood does not appear to have an immediate interpretation via leave-one-out cross-validation, because the data are modeled jointly with an unknown parameter. More work could be done to expand on this point.


The \citeA{zhang2019} models also illustrate that, while the dichotomous ``conditional/marginal'' distinction is most relevant for psychometric models, middle grounds are evident in some situations. Multiple marginal versions of the information criteria exist when the model includes more than two levels and/or (partially-)crossed random effects. Further, in non-multilevel IRT contexts, both the item parameters and person parameters can be modeled as random effects \cite<e.g.,>{deb08}. Here, a researcher might be interested in the model's predictive ability when new people complete the same items that originally entered in the model. This would lead the researcher to compute information criteria where the person random effects are marginalized out of the likelihood but the item random effects are not. These scenarios highlight that, for latent variable models, it is insufficient to report a model information criterion without providing additional detail on how the criterion was computed. Further, the values that are automatically obtained from MCMC software are generally not suitable for further use.

\newpage

\bibliographystyle{apacite}
\bibliography{../refs}

\clearpage
\appendix

\section{Posterior expectations of marginal and conditional likelihoods}
Following \citeA{tregel03}, who studied DIC in the context of linear mixed models, we can use Jensen's inequality to show that the
posterior expected value of the marginal log-likelihood
is less than the posterior expected value of the conditional log-likelihood.

First, consider the function $h(x) = x\log(x)$. It is convex, so Jensen's inequality states that
\begin{equation}
  h(\mathrm{E}(x)) \leq \mathrm{E}(h(x)).
\end{equation}
Setting $x = f_{\text{c}}(\bm{y} | \bm{\omega}, \bm{\zeta})$ and taking expected values with respect to $\bm{\zeta}$, we have that
\begin{align}
  h(\mathrm{E}(x)) &= \displaystyle \log \left [ \displaystyle \int f_{\text{c}}(\bm{y} | \bm{\omega}, \bm{\zeta}) g(\bm{\zeta} | \bm{\psi}) \di \bm{\zeta} \right ] \int f_{\text{c}}(\bm{y} | \bm{\omega}, \bm{\zeta}) g(\bm{\zeta} | \bm{\psi}) \di \bm{\zeta} \\
  \mathrm{E}(h(x)) &= \displaystyle \int \log(f_{\text{c}}(\bm{y} | \bm{\omega}, \bm{\zeta})) f_{\text{c}}(\bm{y} | \bm{\omega}, \bm{\zeta}) g(\bm{\zeta} | \bm{\psi}) \di \bm{\zeta},
\end{align}
so that
\begin{equation}
\displaystyle  \log \left [ \displaystyle \int f_{\text{c}}(\bm{y} | \bm{\omega}, \bm{\zeta}) g(\bm{\zeta} | \bm{\psi}) \di \bm{\zeta} \right ] \int f_{\text{c}}(\bm{y} | \bm{\omega}, \bm{\zeta}) g(\bm{\zeta} | \bm{\psi}) \di \bm{\zeta} \leq \displaystyle \int \log(f_{\text{c}}(\bm{y} | \bm{\omega}, \bm{\zeta})) f_{\text{c}}(\bm{y} | \bm{\omega}, \bm{\zeta}) g(\bm{\zeta} | \bm{\psi}) \di \bm{\zeta}.
\end{equation}

We now multiply both sides of this inequality by $p(\bm{\omega}, \bm{\psi})/c$, where $p(\bm{\omega}, \bm{\psi})$ is a prior distribution and
\begin{equation}
  c = \displaystyle \int_{\bm{\psi}} \displaystyle \int_{\bm{\omega}} \displaystyle \int_{\bm{\zeta}} f_{\text{c}}(\bm{y} | \bm{\omega}, \bm{\zeta}) g(\bm{\zeta} | \bm{\psi}) \di \bm{\zeta} \cdot p(\bm{\omega}, \bm{\psi}) \di \bm{\omega} \di \bm{\psi}
\end{equation}
is the posterior normalizing constant.
Finally, we integrate both sides with respect to $\bm{\omega}$ and $\bm{\psi}$ to obtain
\begin{multline}
\label{eq:ineq}
\displaystyle \int_{\bm{\psi}} \displaystyle \int_{\bm{\omega}} \log \left ( \displaystyle \int_{\bm{\zeta}} f_{\text{c}}(\bm{y} | \bm{\omega}, \bm{\zeta}) g(\bm{\zeta} | \bm{\psi}) \di \bm{\zeta} \right ) \displaystyle \int_{\bm{\zeta}} f_{\text{c}}(\bm{y} | \bm{\omega}, \bm{\zeta}) g(\bm{\zeta} | \bm{\psi}) \di \bm{\zeta} \cdot \left [ p(\bm{\omega}, \bm{\psi})/c \right ] \di \bm{\omega} \di \bm{\psi} \leq \\
\displaystyle \int_{\bm{\psi}} \displaystyle \int_{\bm{\omega}} \displaystyle \int_{\bm{\zeta}} \log \left ( f_{\text{c}}(\bm{y} | \bm{\omega}, \bm{\zeta}) \right ) f_{\text{c}}(\bm{y} | \bm{\omega}, \bm{\zeta}) g(\bm{\zeta} | \bm{\psi}) \di \bm{\zeta} \cdot \left [ p(\bm{\omega}, \bm{\psi})/c \right ] \di \bm{\omega} \di \bm{\psi}.
\end{multline}
We can now recognize both sides of~\eqref{eq:ineq} as expected values of log-likelihoods with respect to the model's posterior distribution, leading to
\begin{equation}
\mathrm{E}_{\bm{\omega}, \bm{\psi} | \bm{y}} \left [ \log f_\tm (\bm{y} |\bm{\omega}, \bm{\psi} ) \right ] \leq \mathrm{E}_{\bm{\omega}, \bm{\zeta} | \bm{y}} \left [ \log f_\tc (\bm{y}|\bm{\omega}, \bm{\zeta}) \right ].
\end{equation}
Note that the above results do not rely on normality, so they also apply to, e.g., the two-parameter logistic model estimated via marginal likelihood.


\section{Effective number of parameters for marginal and conditional DIC}
To consider the effective number of parameters for normal likelihoods, we rely on results from
\citeA{spibes02}. They showed that the effective number of parameters $p_D$ can be viewed as the fraction of information about model parameters in the likelihood, relative to the total information contained in both the likelihood and prior. Under this view, a specific model parameter gets a value of ``1'' if all of its information is contained in the likelihood, and it gets a value below ``1'' if some information is contained in the prior. We sum these values across all parameters to obtain $p_D$.

\citeA{spibes02} relatedly showed that, for normal likelihoods, $p_D$ can be approximated by
\begin{equation}
p_D \approx \text{tr}(\bm{I}(\hat{\bm{\theta}}) \bm{V}),
\end{equation}
where $\bm{\theta}$ includes all model parameters (including $\bm{\zeta}$ in the conditional model), $\bm{I}(\hat{\bm{\theta}})$ is the observed Fisher information matrix, and $\bm{V}$ is the posterior covariance matrix of $\bm{\theta}$. When the prior distribution of $\bm{\theta}$ is noninformative, then $\bm{I}(\hat{\bm{\theta}}) \approx \bm{V}^{-1}$. Consequently, matching the discussion in the previous paragraph, the effective number of parameters under noninformative priors will approximate the total number of model parameters.

This result implies that the conditional $p_D$ will tend to be much larger than the marginal $p_D$. In particular, in the conditional case, each individual has a unique $\bm{\zeta}_j$ vector that is included as part of the total parameter count. The resulting $p_D$ will not necessarily be close to the total parameter count because the ``prior distribution'' of $\bm{\zeta}_j$ is a hyperdistribution, whereby individuals' $\bm{\zeta}_j$ estimates are shrunk towards the mean. Thus, for these parameters, the ``prior'' is informative. However, even when the fraction of information in the likelihood is low for these parameters, the fact that we are summing over hundreds or thousands of $\bm{\zeta}_j$ vectors implies that the conditional $p_D$ will be larger than the marginal $p_D$.

\section{Adaptive Gaussian quadrature for marginal likelihoods}

 We modify the adaptive quadrature method proposed by \citeA{rabe2005maximum} for generalized linear mixed models to
 a form designed to exploit MCMC draws from the joint posterior of all latent variables and model parameters. Here we describe one-dimensional integration, but the method is straightforward to generalize to multidimensional integration as in \citeA{rabe2005maximum}. We assume that $\zeta_j$ represents a disturbance with zero mean and variance $\tau^2$ so that there is only one hyperparameter $\psi=\tau$.

In a non-Bayesian setting,
standard (non-adaptive) Gauss-Hermite quadrature can be viewed as approximating the conditional prior density $g(\zeta_j|\tau)$  by a discrete distribution with masses $w_m$, $m=1,\ldots, M$ at locations $a_m\tau$ so that
the integrals in~(\ref{eq:likem}) are approximated by sums of $M$ terms, where $M$ is the number of quadrature points,
\begin{equation}
 f_\tm(\bm{y}| \bm{\omega}, {\tau})\ \approx \   \prod_{j=1}^J \sum_{m=1}^M w_m f_\tc(\bm{y}_{j} | \bm{\omega}, \zeta_j=a_m\tau) .
 \label{eq:likemad}
\end{equation}
To obtain information criteria, this method can easily be applied to the conditional likelihood function for each draw $\bm{\omega}^s$ and $\tau^s$ ($s=1,\ldots,S$) of the model parameters from MCMC output.
This approximation can be thought of as a deterministic version of Monte Carlo integration. Adaptive quadrature is then a deterministic version of Monte Carlo integration via importance sampling.

If applied to each MCMC draw $\bm{\omega}^s$ and $\tau^s$, the importance density is a normal approximation to the \textit{conditional} posterior density of $\zeta_j$, given the current draws of the model parameters. \citeA{rabe2005maximum} used a normal density with  mean and variance equal to the mean $\E(\zeta_j|\bm{y}_j, \bm{\omega}^s, \tau^s)$ and variance $\mathrm{var}(\zeta_j|\bm{y}_j, \bm{\omega}^s, \tau^s)$ of the
conditional posterior density of $\zeta_j$, whereas \citeA{pinheiro:95} and some software use a normal
density with mean equal to the mode of the conditional posterior and variance equal to minus the reciprocal of the second derivative of the conditional log posterior.
Here, we modify the method by \citeA{rabe2005maximum} for the Bayesian setting by using a normal approximation to the \textit{unconditional} posterior density of $\zeta_j$ as importance density.
Specifically, we use a normal density with mean
\begin{equation}
  \tilde \mu_j =
  \widetilde{E}(\zeta_j | \bm{y}) =
  \frac{1}{S} \sum_{s=1}^S \zeta_j^s
,\end{equation}
and standard deviation
\begin{equation}
  \tilde \phi_j =
  \sqrt{\widetilde\mathrm{var}(\zeta_j | \bm{y}) } =
  \sqrt{\frac{1}{S-1} \sum_{s=1}^S (\zeta_j^s- \tilde \mu_j)^2},
  \end{equation}
  where $\zeta_j^s$ is the draw of $\zeta_j$ from its unconditional posterior in the $s$th MCMC iteration.
  The tildes indicate that these quantities are subject to Monte Carlo error.

  Note that this version of adaptive quadrature is computationally more efficient than the one based on the mean and standard deviation of the \textit{conditional} posterior distributions because the latter would have to be evaluated for each MCMC draw and would require numerical integration, necessitating a procedure that iterates between updating the quadrature locations and weights and updating the conditional posterior means and standard deviations. A disadvantage of our approach is
  that the quadrature locations and weights (and hence importance density) are not as targeted, but the computational efficiency gained also makes it more feasible to increase $M$.

The adaptive quadrature approximation to the marginal likelihood for cluster $j$ at posterior draw $s$ becomes
\begin{equation}
 f_\tm(\bm{y}| \bm{\omega}, {\tau})\ \approx \   \prod_{j=1}^J
  \sum_{m=1}^M
    w_{jm}^s f_\tc(\bm{y}_{j} | \bm{\omega}, \zeta_j=a_{jm})
,\end{equation}
where the adapted locations are
\begin{equation}
  a_{jm} = \tilde \mu_j +
                    \tilde \phi_j \times
                    a_m
,\end{equation}
and the corresponding weights or masses are
\begin{equation}
  w_{jm}^s = \sqrt{2\pi} \times
        \tilde \phi_j \times
        \exp \left ( \frac{a_{m}^2}{2} \right ) \times
        g \left ( a_{jm}; 0, \tau^{2,s} \right ) \times
        w_m,\end{equation}
where $ g \left ( a_{jm}; 0, \tau^{2,s} \right )$ is the normal density function with mean zero and variance $\tau^{2,s}$, evaluated at $a_{jm}$.

The number of integration points $M$ required to obtain a sufficiently accurate approximation is determined by evaluating the approximation of the target quantity (DIC, WAIC) with increasing values of $M$ (7, 11, 17, etc.) and choosing the value of $M$ for which the target quantity changes by less than 0.01  from the previous value. Here the candidate values for $M$ are chosen to increase
approximately by 50\% while remaining odd so that one of the quadrature locations is at the posterior mean.
\citeA{furr2017} finds this approach to be accurate in simulations for linear mixed models where the adaptive quadrature approximation can be
compared with the closed form integrals.

\section{Monte Carlo Error for the DIC and WAIC effective number of parameters}

For the DIC effective number of parameters, $p_\D$, we can make use of the well-known method for estimating the Monte-Carlo error for the mean of a quantity across MCMC iterations.
Let a quantity computed in  MCMC iteration $s$ ($s=1,\ldots,S$) be denoted $\gamma_s$, so the point estimate of the expectation of $\gamma$ is
\[
\overline{\gamma} \ = \ \frac{1}{S}\sum_{s=1}^S \gamma_s.
\]
Then the squared Monte Carlo error (or Monte Carlo error variance) is estimated as
\begin{equation}
{\rm MCerr}^2(\overline{\gamma})\ =\ \frac{1}{S_{\rm{}eff}}\left[ \frac{1}{S-1} \sum_{s=1}^S (\gamma_s-\overline{\gamma})^2\right],
\label{eq:errmean}
\end{equation}
where $S_{\rm{}eff}$ is the effective sample size.

For the effective number of parameter approximation in~(\ref{eq:plumopt}) proposed by Plummer (2008), we can obtain the Monte Carlo error variance by substituting
\[ \gamma_s\ =\ \frac{1}{2} \log \left \{ \frac{f(\bm{y}_s^{\R1} | \bm{\theta}_s^1)}{f(\bm{y}_s^{\R1} | \bm{\theta}_s^2)} \right \} + \frac{1}{2} \log \left \{ \frac{f(\bm{y}_s^{\R2} | \bm{\theta}_s^2)}{f(\bm{y}_s^{\R2} | \bm{\theta}_s^1)} \right \}
\]
in~(\ref{eq:errmean}).

For the effective number of parameter approximation in~(\ref{eq:pd}) proposed by Spiegelhalter et al. (2002), we assume that the variation due to $\E_{\bm{\theta}|\bm{y}}[-2\log f(\bm{y}|{\bm{\theta}})]$ dominates and use
\[
\gamma_s = -2\log f(\bm{y}|{\bm{\theta}_s}).
\]

For the WAIC effective number of parameters, $p_\W$, we use expressions for the Monte Carlo error of sample variances \cite<see, e.g.>{white10}. Let the variance
of $\gamma_s$ over MCMC iterations be denoted $v(\gamma)$,
\[
v(\gamma)\ = \ \frac{1}{S-1} \sum_{s=1}^S (\gamma_s-\overline{\gamma})^2 = \frac{1}{S} \sum_{s=1}^S T_s,\quad\quad T_s = \frac{S}{S-1}(\gamma_s-\overline{\gamma})^2
\]
Then the Monte Carlo error variance is estimated as
\begin{equation}
{\rm MCerr}^2(v(\gamma))\ =\ \frac{1}{S_{\rm{}eff}\times S} \sum_{s=1}^S (T_s-v(\gamma))^2,
\label{eq:errvar}
\end{equation}

The conditional version of the effective number of parameters is given by the sum over all units of the posterior variances of the pointwise log posterior densities,
 \[
 p_{\W\tc} \ = \ \sum_{j=1}^J\sum_{i=1}^{n_j}  \var_{\bm{\omega},\bm{\zeta}|\bm{y}} \left[ \log f_\tc({y}_{ij}|\bm{\omega},\bm{\zeta}_j)\right].
 \]
The posterior variance $\var_{\bm{\omega},\bm{\zeta}|\bm{y}} \left[ \log f_\tc({y}_{ij}|\bm{\omega},\bm{\zeta}_j)\right]$ for a given unit
is estimated by $v(\gamma_{ij})$ with
\[ \gamma_{ijs}\ =\  \log f_\tc({y}_{ij}|\bm{\omega}_s,\bm{\zeta}_{js}), \]
where we have added subscripts $ij$ to identify the unit, and has  Monte Carlo error variance ${\rm MCerr}^2(v(\gamma_{ij}))$ given in~(\ref{eq:errvar}). The variance  of the sum of the independent contributions $v(\gamma_{ij})$ to $\hat{p}_{\W\tc}$ is the sum of the variances of these contributions,
\[
{\rm MCerr}^2(\hat{p}_{\W\tc})\ = \  \sum_{j=1}^J\sum_{i=1}^{n_j} {\rm MCerr}^2(v(\gamma_{ij})).
\]
 For the marginal version of the effective number of parameters, $p_{\W\tm}$, we define
 \[
 \gamma_{js}\ =\  \log f_\tm(\bm{y}_{j}|\bm{\omega}_s,\bm{\psi}_s)
 \]
 and
 \[
{\rm MCerr}^2(\hat{p}_{\W\tm})\ = \  \sum_{j=1}^J{\rm MCerr}^2(v(\gamma_{j})).
\]

\section{Additional results}

This section contains additional results from the CFA example that were not included in the main text.

\begin{figure}
\caption{Marginal and conditional DICs (Spiegelhalter et al.\ definitions) under uninformative prior distributions for nine models from Wicherts et al.}
\label{fig:wicdicwk}
\begin{center}
\includegraphics[width=5in]{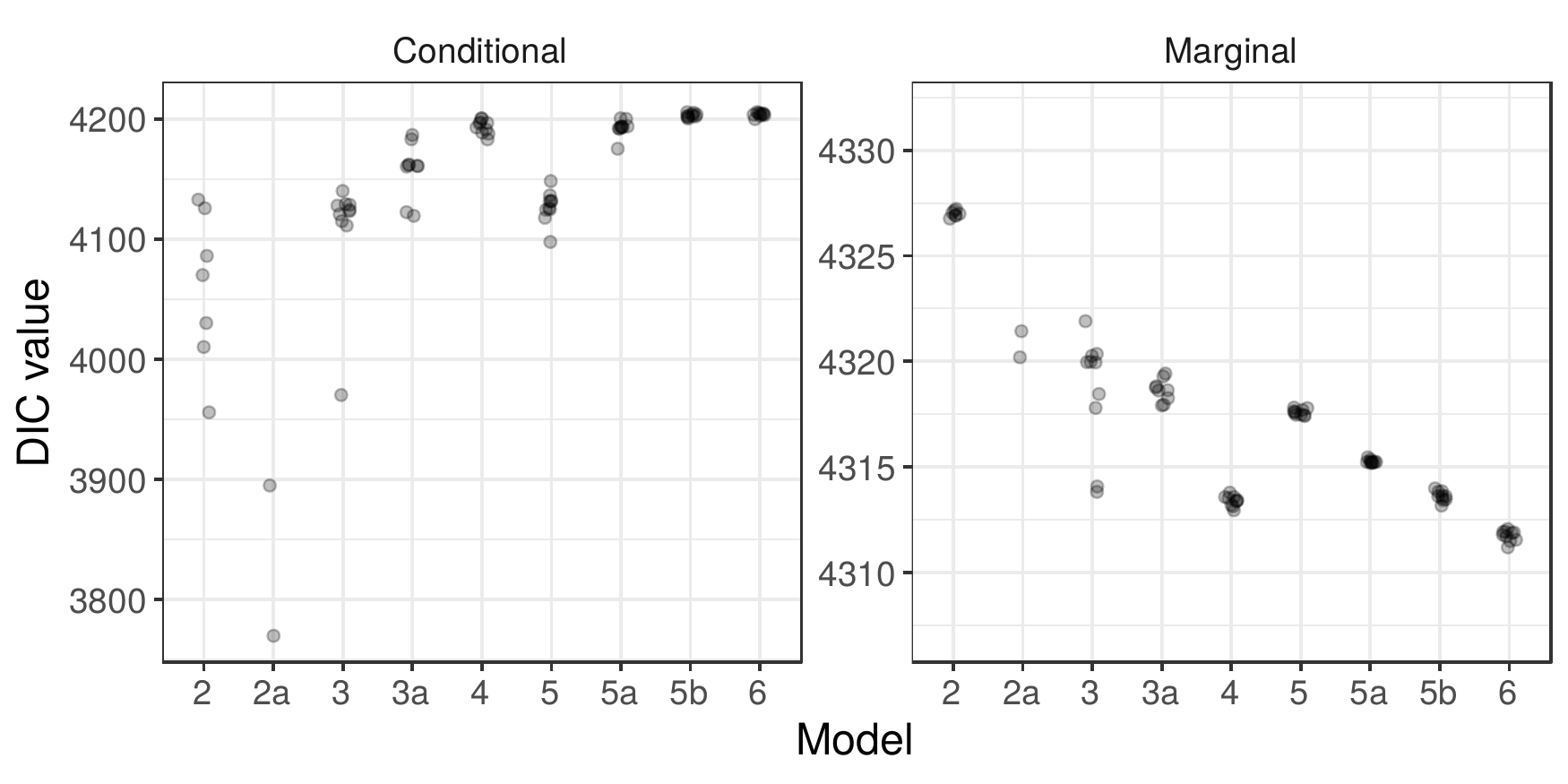}
\end{center}
\end{figure}

Figure~\ref{fig:wicdicwk} shows Spiegelhalter DIC values for models that use the uninformative priors described in the ``Prior sensitivity'' subsection. Of note here is that Models 2 and 2a sometimes failed to converge, resulting in fewer than ten points in the graphs. Because we used the automatic convergence procedure described in the main text, ``failure to converge'' here means that the chains did not achieve Gelman-Rubin statistics below 1.05 in the five minutes allotted. When we removed the 5-minute maximum time to convergence, we encountered situations where chains ran for days without converging. In our experience, these convergence issues are often observed for CFA models in JAGS with flat priors. Chains sometimes get stuck in extreme values of the parameter space and cannot recover.

\begin{figure}
\caption{Marginal and conditional DICs (Plummer definitions) under informative prior distributions for nine models from Wicherts et al.}
\label{fig:wicdicjag}
\begin{center}
\includegraphics[width=5in]{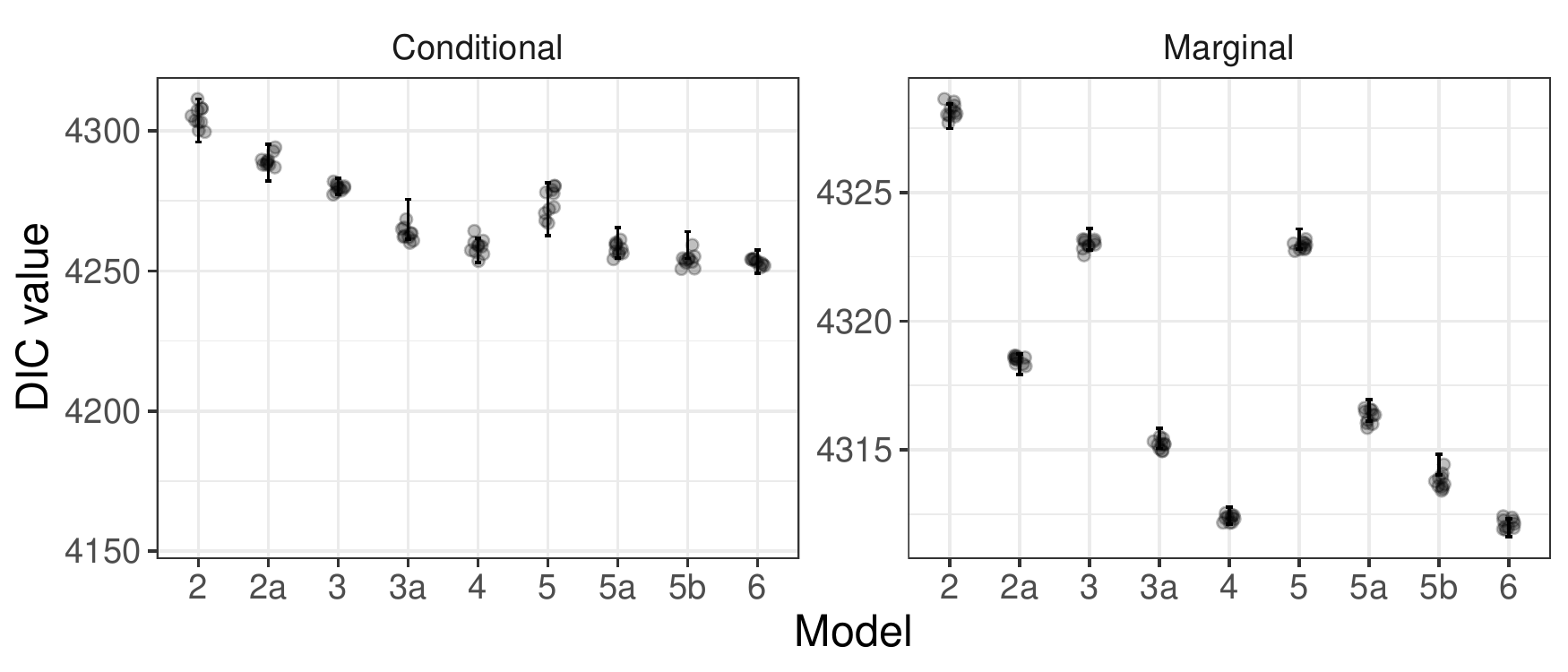}
\end{center}
\end{figure}

Figure~\ref{fig:wicdicjag} shows Plummer DIC values for models that use the informative priors described in the ``Prior sensitivity'' subsection. The figure also contains error bars ($\pm 2$ SDs) from a single replication, similarly to Figure~\ref{fig:jagerr} in the main text. These error bars appear to continue to track Monte Carlo error in DIC. Comparing Figure~\ref{fig:wicdicjag} to Figure~\ref{fig:wicdicstr}, we observe a different pattern in the conditional DICs across the Plummer and Spiegelhalter definitions. The Spiegelhalter conditional DICs (Figure~\ref{fig:wicdicstr}) consistently prefer Model 2a, whereas the Plummer conditional DICs (Figure~\ref{fig:wicdicjag}) generally decrease across models and become lowest for the final models, labeled 5b and 6 (though Models 4 and 5a are also similar). On the other hand, the marginal DICs are similar across the Spiegelhalter and Plummer definitions.

\end{document}